\documentclass[]{bytedance_seed}

\usepackage[toc,page,header]{appendix}

\usepackage{minitoc}
\usepackage{algorithm}
\usepackage{algpseudocode}
\usepackage{xcolor}
\usepackage{soul}
\usepackage{multicol}
\usepackage{multirow}
\usepackage{pifont}

\definecolor{darkgreenRGB}{RGB}{0, 128, 0}

\title{ Triton-distributed: Programming Overlapping Kernels on Distributed AI Systems with the Triton Compiler }

\author[1,2, \dagger]{Size Zheng}
\author[1, \dagger]{Wenlei Bao}
\author[1]{Qi Hou}
\author[1]{Xuegui Zheng}
\author[1]{Jin Fang}
\author[1]{Chenhui Huang}
\author[1, 3]{Tianqi Li}
\author[1, 4]{Haojie Duanmu}
\author[1, 3]{Renze Chen}
\author[1, 3]{Ruifan Xu}
\author[1, 5]{Yifan Guo}
\author[1]{Ningxin Zheng}
\author[1]{Ziheng Jiang}
\author[1]{Xinyi Di}
\author[1]{Dongyang Wang}
\author[1]{Jianxi Ye}
\author[1]{Haibin Lin}
\author[1]{Li-Wen~Chang}
\author[5]{Liqiang Lu}
\author[3]{Yun Liang}
\author[2]{Jidong Zhai}
\author[1, \dagger]{Xin Liu}

\affiliation[1]{ByteDance Seed}
\affiliation[2]{Tsinghua University}
\affiliation[3]{Peking University}
\affiliation[4]{Shanghai Jiao Tong University}
\affiliation[5]{Zhejiang University}

\contribution[\dagger]{Corresponding authors}

\abstract{
As the scaling of single chip is gradually approaching its bottleneck, a single accelerator can no longer support the training and inference of existing large language models. Therefore, it has become a pressing need to use distributed system composed of multiple accelerators for training and inference. In a distributed system, there are three fundamental activities occur concurrently: computation, memory access, and communication. 
In existing training/inference frameworks, these aspects are often optimized independently at different programming levels. As a result, it is difficult for these activities to coordinate with each other and unleash the full performance  potential of the cluster.

In this report, we propose Triton-distributed, an extension of existing Triton compiler, to overcome the programming challenges in distributed AI systems.
Triton-distributed is the first compiler that supports native overlapping optimizations for distributed AI workloads, providing a good coverage of existing optimizations from different frameworks.
First, we integrate communication primitives compliant with the OpenSHMEM standard into the compiler. This enables programmers to utilize these primitives with a higher-level Python programming model.
Second, we illustrate how to achieve complex joint optimization of computation, memory access, and communication with the assistance of the compiler. In particular, we show how to use overlapping techniques to hide latency and present our compiler-based programming methods in both single-node and multi-node scenarios.
Finally, we showcase the performance of the code generated by our compiler. In a test environment with up to 64 devices, our compiler can fully utilize heterogeneous communication and computation resources to provide effective overlapping and high performance. In many cases, the performance of the generated code can even outperform hand-optimized code. Moreover, the development difficulty and the time cost for development using our compiler are far less than those of low-level programming such as CUDA/C++, which clearly demonstrates significant productivity advantages. 
}

\date{\today}
\correspondence{Size Zheng at \email{zheng.size@bytedance.com}, Wenlei Bao at \email{wenlei.bao@bytedance.com}, Xin Liu at \email{liuxin.ai@bytedance.com}}

\checkdata[Project Page]{\url{https://github.com/ByteDance-Seed/Triton-distributed}}

\begin{document}
\maketitle

\newpage
\tableofcontents
\newpage

\section{Introduction}
With the rapid development of AI, leveraging extremely large language models (e.g., ChatGPT~\cite{chatgpt}, Qwen~\cite{qwen-max}, DeepSeek~\cite{deepseek-v3}, Doubao~\cite{doubao-1.5}, etc.), remarkable progress far exceeding expectations has been made in various fields, including chat~\cite{chatgpt, doubao-1.5}, writing, question answering, coding, mathematics~\cite{qwen-max, deepseek-v3}, image generation~\cite{4o-image}, and video generation~\cite{sora}. The efficient deployment of AI models depends on the co-optimization of underlying hardware and software. 
At the software level, the main task is to map large-scale computations and parameters onto hardware. 
As models grow larger, the target hardware for mapping has shifted from single device to multi-accelerator systems.


Distributed programming is notoriously difficult. Despite the extensive work on distributed systems over the past few decades \cite{gpipe, megatron-lm, deepspeed, overlap-1, overlap-2, overlap-3, overlap-4}, these efforts either address only the issues in CPU clusters or require extremely complex engineering optimizations to achieve good performance on AI accelerators.
Moreover, there is a significant gap between distributed development and AI algorithm development. Distributed development generally requires programming in CUDA/C++, while algorithm development is often carried out in Python. This inherently necessitates cross-language programming. As a result, most users are proficient in optimizing only one of these two aspects. To achieve both algorithm development and underlying optimization simultaneously, one needs to bridge a significant programming gap. It usually also requires cross-team collaboration, which further leads to a decrease in development efficiency.


Over the past decades, the development of high-performance operators using Python programming has emerged as a key research area, attracting extensive exploration from both academia and industry. This process led to the emergence of numerous compilers~\cite{halide, tvm, triton, mlir, flextensor, amos, ansor, pallas, tensorir, taco, taso}, but after years of validation, the outstanding compilers that have withstood the test of time~\cite{triton, tilelang, mlir} still mainly focus on single-device scenarios.
In terms of single chip code generation capability, these compilers have already matched the level of experts. For instance, on NVIDIA GPUs, their performance rivals that of CUTLASS~\cite{cutlass} and cuBLAS~\cite{cublas}. Some compilers, such as Triton~\cite{triton}, have been adapted by numerous manufacturers to support a wide range of AI chips~\cite{amd-fused, triton-mlu}.
However, with the advent of LLMs, research on compiler optimization for single chip has almost converged (although hardware advancements continue to drive the development of new compilation techniques). Overall, compiler research has entered a distributed era. Early studies on distributed compilers, such as distributed Halide~\cite{dist-halide} and DISTAL~\cite{distal}, were confined to non-large-model scenarios and fell short of meeting the distributed requirements of emerging LLMs. Additionally, previous distributed compilers were inclined to propose proprietary DSLs rather than align with Python.

\begin{figure}[t]
    \centering
\includegraphics[width=0.9\textwidth]{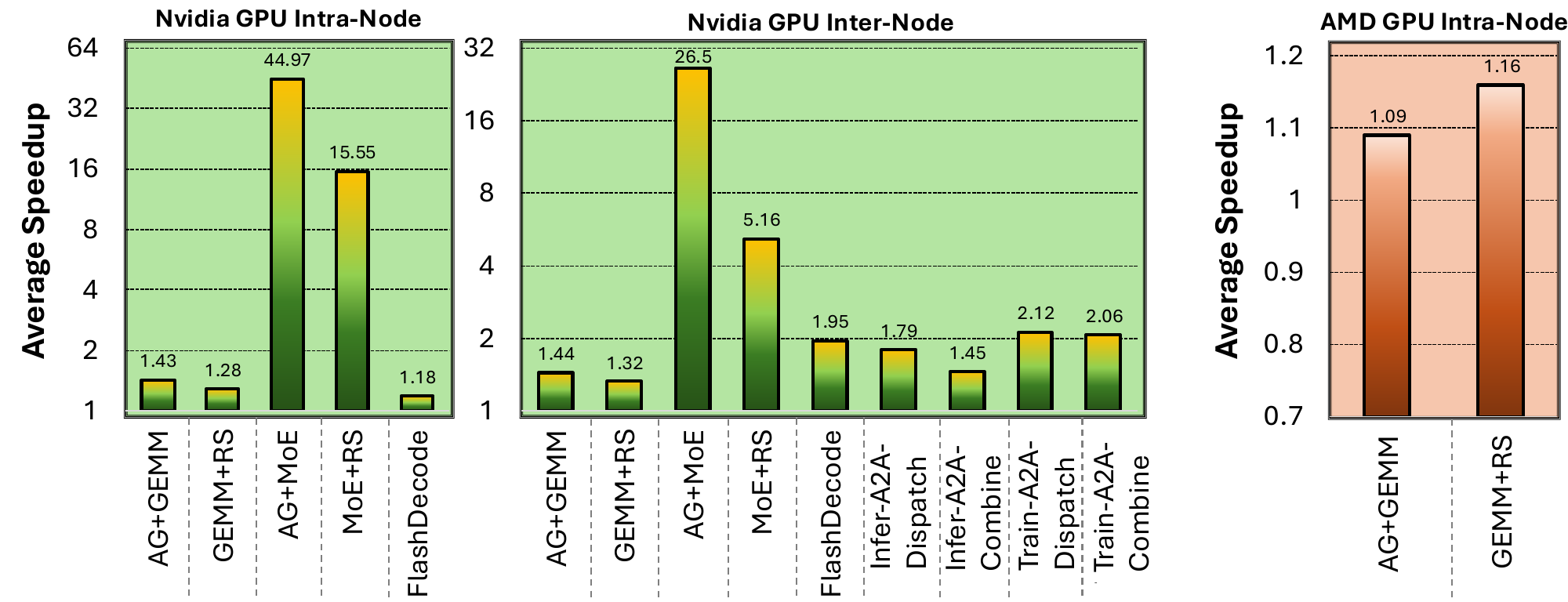}
    \caption{Average Speedup of Triton-distributed to Baselines (PyTorch+NCCL/RCCL).}
    \label{fig:overall-perf}
\end{figure}

For LLMs, the key requirement for distributed optimization becomes computation-communication overlapping. Previously, in small-scale distributed training/inference, communication overhead wasn’t a critical cost issue. However as cluster number scales exponentially, overlapping computation with communication becomes vital. For example, ByteDance's COMET~\cite{comet} recently saved millions of GPU hours through this technique, equivalent to millions of dollars in cost savings.
The ability to overlap computation with communication has exceeded the scope of existing compilers, leaving this optimization accessible only to a few teams with exceptional engineering capabilities~\cite{flux, deepep, centauri}.
In this report, we propose Triton-distributed, an distributed compiler extension based on open-source Triton. Triton-distributed supports native fine-grained computation-communication overlapping using compiler-assisted primitives. Both computation implementation and communication optimization are fully achieved at the Python level, yielding performance comparable to or better than CUDA/C++ implementations.
In Figure~\ref{fig:overall-perf}, we show the overall speedup of Triton-distributed to NCCL/RCCL on Nvidia GPUs and AMD GPUs for a wide range of workloads.
The speedup ranges from $1.09\times$ to $44.97\times$.
Trition-distributed supports both overlapping tasks such as AllGather GEMM (AG+GEMM) and GEMM ReduceScatter (GEMM+RS) as well as communication-only tasks such as expert-parallel AllToAll (inference low-latency AllToAll dispatch/combine and training high-bandwidth AllToAll dispatch/combine).
All these workloads are implemented using our compiler-assisted primitives.
Our compiler translates these primitives into NVSHMEM/ROCSHMEM implementations during compilation, enabling communication across both single-node and multi-node GPU clusters.
Notably, implementing these workloads requires minimal modifications to Triton's original compute code: developers only focus on adding communication logic.

\begin{figure}[t]
    \centering
\includegraphics[width=0.8\textwidth]{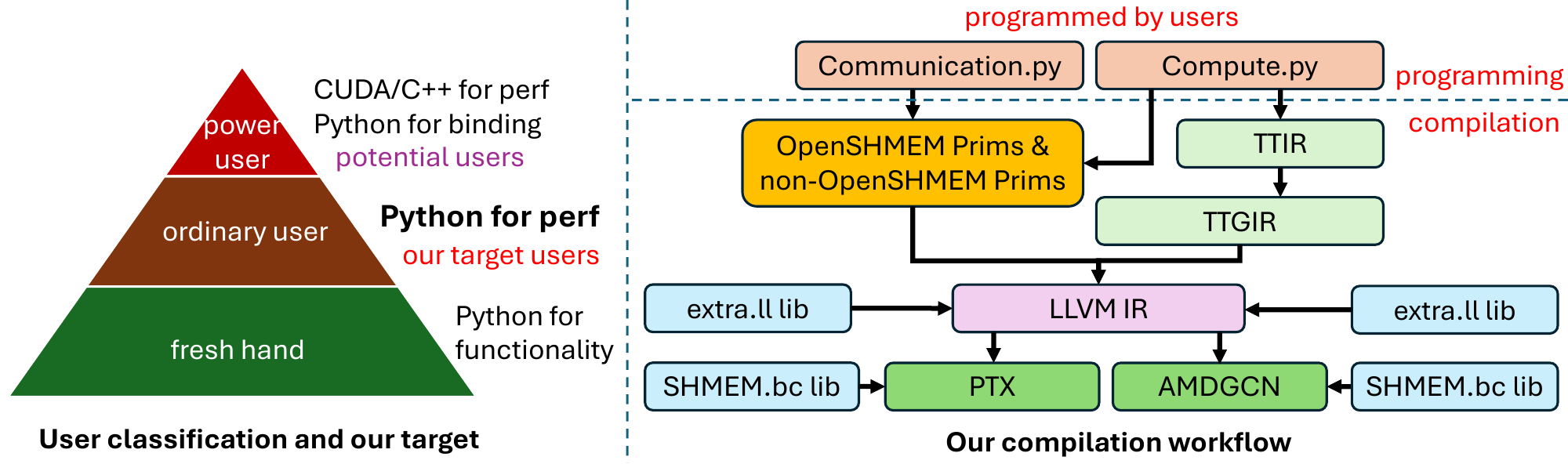}
    \caption{Overview of our compilation stack.}
    \label{fig:intro}
\end{figure}

In Figure~\ref{fig:intro}, we present an overview of our compilation stack. Users can program communication and computation parts separately in Python. The communication part is implemented with OpenSHMEM primitives, which are further lowered to LLVM IR along with a bitcode library. The computation part goes through the original Triton compilation stack to LLVM IR. Finally, PTX or AMDGCN are generated and executed on target accelerators.
Our compiler is mainly designed for ordinary users who are familiar with Python-level LLM model development and have an average knowledge of performance optimization. Power users who are experts in low-level programming may also use our compiler to reduce development overhead without much performance degradation. 
Furthermore, leveraging Triton's multi-hardware support, our solution also works on AMD GPUs and other NPUs. Based on our compiler, we have also built a higher-level compiler TileLink~\cite{tilelink} with higher-level communication primitives (which are not included in this report).
\section{Programming Model}
To use our compiler, we first introduce the programming model. Then we introduce the primitives supported by our compiler. Finally, we show an AllGather GEMM example written using our programming model.

\begin{table}[h]
    \centering
    \caption{Our Communication Primitives}
    \label{table:primitives}
    \begin{tabular}{ll}
        \toprule
        \textbf{Primitive} & \textbf{Explanation}   \\
        \midrule
        & \textbf{OpenSHMEM Primitives}\\
        \midrule
        \textit{my\_pe} & Get the current device id  \\
        \textit{n\_pes} & The number of devices in the world  \\
        \textit{int\_p} & Put an integer to remote device  \\
        \textit{remote\_ptr} & Convert local shared memory pointer to remote pointer  \\
        \textit{barrier\_all} & Barrier all the devices   \\
        \textit{sync\_all} & Synchronize all the devices  \\
        \textit{quiet} & Ensure completion of shared memory operation of calling device  \\
        \textit{fence} & Ensure order of shared memory operation of calling device  \\
        \textit{getmem} & Blocking get data from remote device  \\
        \textit{getmem\_nbi} & Non-blocking get data from remote device  \\
        \textit{putmem} & Blocking put data to remote device  \\
        \textit{putmem\_nbi} & Non-blocking put data to remote device  \\
        \textit{putmem\_signal} & Blocking put data and write signal to remote device  \\
        \textit{putmem\_signal\_nbi} & Non-blocking put data and write signal to remote device  \\
        \textit{signal\_op} & Perform signal set/add operation to remote  \\
        \textit{signal\_wait\_until} & Wait local signal until condition is meet  \\

        \textit{broadcast} & Broadcast data into all the other ranks \\

        \midrule
        & \textbf{non-OpenSHMEM Primitives}  \\
        \midrule

        \textit{wait} & Locally wait a signal until it equals to some given value  \\
        \textit{consume\_token} & used with \textit{wait} primitive to create data dependency \\
        \textit{notify} & Notify a remote signal, similar to \textit{signal\_op} primitive \\

        \textit{atomic\_cas} & Atomic compare and swap  \\

        \textit{atomic\_add} & Atomic add value  \\

        \textit{ld\_acquire} & Load with acquire semantic  \\

        \textit{red\_release} & Reduction add with release semantic  \\

        \textit{multimem\_ld\_reduce} & Multimem load data and perform reduction  \\

        \textit{multimem\_st} & Multimem broadcast data  \\

        \bottomrule
    \end{tabular}
\end{table}

\subsection{MPMD Programming: Symmetric Memory, Signal Exchange, and Async-Task}
Our programming model follows the MPMD (multiple programs multiple data) model. The MPMD model allows communication and computation tasks to run in parallel and cooperate with each other to complete a global task. The core of our programming model includes three concepts: symmetric memory, signal exchange, and async-task.

\textbf{Symmetric Memory:} Each rank allocates a memory buffer in the global scope with the same size. Each memory buffer has a separate address space, and there is no uniform virtual address (UVA) space from a global perspective. Remote memory buffers cannot be accessed directly via pointers. To perform remote data transfer, specific primitives are required.

\textbf{Signal Exchange:} Operations on each rank use signals to communicate with each other in a consistent manner. A signal is a data object that resides in symmetric memory. There is a fixed set of operations to manipulate signals, including setting the value of a signal, increasing the value of a signal, checking the value of a signal, and performing a spin-lock on a given signal.

\textbf{Async-Task:} Operations such as data transfer and computation are treated as asynchronous tasks that run in parallel. Async-tasks can be synchronized through signals. Note that even on the same rank, the operations are asynchronous. For different hardware backends, there are different ways to implement async-tasks. For GPUs, multi-streaming and multi-threading are common choices. Multi-streaming relies on runtime task queues to launch different tasks simultaneously, while multi-threading leverages parallel hardware units.

\begin{figure}[t]
    \centering
\includegraphics[width=0.8\textwidth]{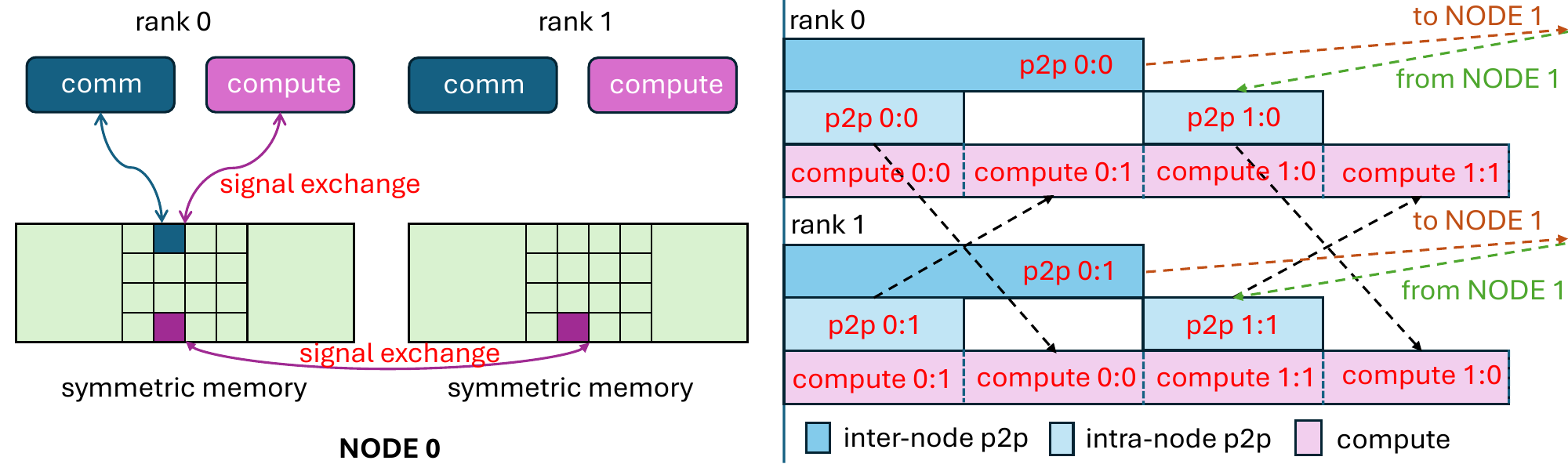}
    \caption{Explanation of Symmetric Memory, Signal Exchange, and Async-Task.}
    \label{fig:programming-model}
\end{figure}

For the example in Figure~\ref{fig:programming-model}, we visually demonstrate the three core concepts. In this example, we use 2 nodes, each with 2 ranks. The left part of Figure~\ref{fig:programming-model} shows the symmetric memory buffers of each rank and the signal exchange between different tasks within the same rank or across ranks.
In the right part, we use a timeline to show that at each rank, three tasks run in parallel: inter-node P2P (point-to-point) data transfer, intra-node P2P data transfer, and computation. Tasks at different ranks also run in parallel. Each rank launches the three tasks simultaneously. The computation task is a single kernel that runs on the device; the order of the tiles of the computation kernel is carefully tuned so that the computation never waits for the communication (note that the computation order at rank 0 and rank 1 is different). In this example, we use \textit{0:0} to denote the data tile from node 0, rank 0. Other notations are similar.

\subsection{Communication Primitives}
For a distributed system, the essence of designing primitives is to model communication patterns effectively. The design of communication primitives depends on the system's architecture, such as the interconnection topology and bandwidth. Currently, mainstream distributed systems like Nvidia GPUs and AMD GPUs support the OpenSHMEM standard and have implemented their own shared-memory communication primitive libraries. Although other NPU accelerators do not currently support the OpenSHMEM standard, it can be expected that this standard will be widely adopted in the future. Therefore, the new primitives we added to the compiler should also align with the OpenSHMEM standard.

Currently, we provide two sets of primitives: OpenSHMEM primitives and non-OpenSHMEM primitives. We list these two sets of primitives in Table~\ref{table:primitives}. For OpenSHMEM primitives, their corresponding implementations can be found in NVSHMEM and ROCSHMEM for Nvidia and AMD GPUs respectively. Non-OpenSHMEM primitives (e.g. \textit{wait}, \textit{consume\_token}, and \textit{notify}) provide complementary functions. These primitives are specially designed for optimization purposes. For example, \textit{wait} is used with \textit{consume\_token} to construct data dependency between signal operations, and the following MMA operations for better compiler-based pipelining. Load/store primitives with specific semantics are used for low-latency communication or signal exchange within a node. These primitives include atomic operations, load with acquire semantics, store with release semantics, and load/store with multimem semantics.


\subsection{Example: Inter-node Overlapping AllGather GEMM}

\begin{figure}[t]
    \centering
\includegraphics[width=0.9\textwidth]{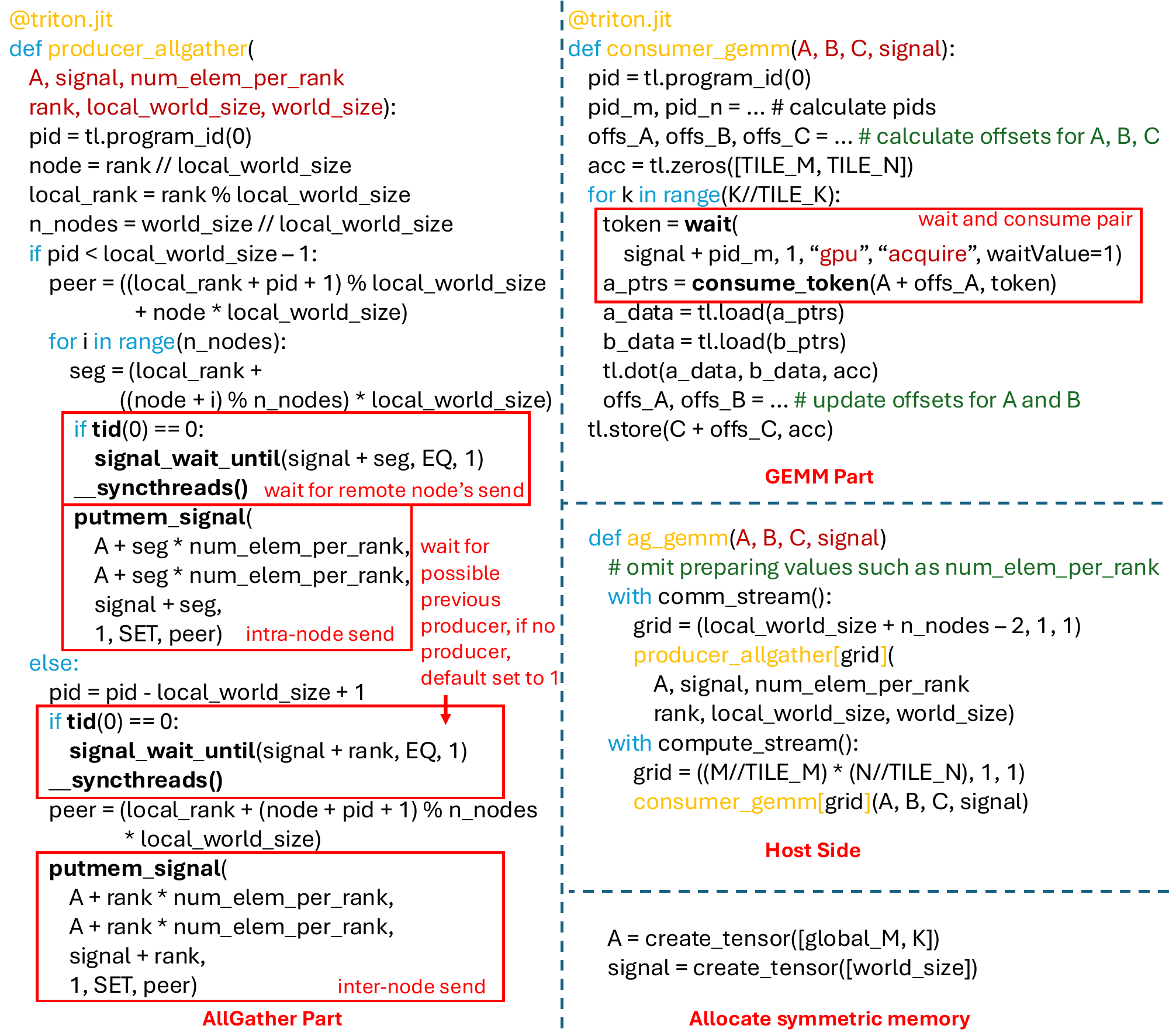}
    \caption{Code Example of AllGather GEMM for Inter-node}
    \label{fig:ag-gemm-example-inter-node}
\end{figure}

In Figure~\ref{fig:ag-gemm-example-inter-node} we show how to program an inter-node overlapping AllGather GEMM using our compiler. The AllGather GEMM program is composed of three parts: the communication part, the computation part, and the host side.
For the communication part (on the left), we assign different tasks to different threadblocks. Part of the threadblocks are responsible for intra-node dispatch, while the other threadblocks perform inter-node data transfer. These two groups of threadblocks run in parallel to overlap inter-node data transfer and intra-node data transfer.
For the computation part (up-right), we just reuse Triton's original GEMM implementation and add only two primitives in the GEMM kernel. The first primitive \textit{wait} produces a token related to a signal, while the second primitive \textit{consume\_token} consumes the token and creates data dependency between the token and the following data load. Different tiles in the GEMM kernel runs in parallel, each tile waits for its own signal, overlapping its dependent communication and other tiles' computation.
Finally, the host-side code (bottom-right) allocates symmetric memory and launches the communication part and the computation part on different streams. The communication part and the computation part both requires SMs (streaming multiprocessors) for execution. 
The execution timeline of this AllGather GEMM is demonstrated in Figure~\ref{fig:programming-model}.
\section{Overlapping Optimizations in Triton-distributed}

Using Triton-distributed, we can cover common overlapping optimizations for distributed workloads.
In this section, we first summarize common overlapping optimizations.
Then, we show kernel implementations using our compiler. To demonstrate the generality of our compiler, we show the kernel optimizations on two platforms: Nvidia GPUs and AMD GPUs.
For kernel implementation, we show the one-sided equivalent of collective communication and how to optimize them for different purposes (high-bandwidth or low-latency). The term one-sided means that all communication operations are programmed from the perspective of a single rank, which is different from collective communication programming, where communication is programmed against all ranks. 
After this, we show how to overlap the communication kernels with computation kernels using signals. To achieve the best performance, tile swizzling is required. Tile swizzling changes the order of communication and computation, making it possible to achieve the maximum overlap in distributed systems. 
Finally, we explain our performance tuning techniques, including automatic distributed program tuning and manual analytical configurations.

\subsection{Optimization Approaches for Overlapping}

\begin{table}[h]
    \centering
    \footnotesize
    \caption{Optimization Approaches and Comparison with Other Frameworks}
    \label{table:optimization-support}
    \begin{tabular}{cccccccc|c}
        \toprule
        \textbf{Name} & \textbf{NCCL} & \textbf{PyTorch} & \textbf{TE} & \textbf{Pallas} & \textbf{CoCoNet} & \textbf{FLUX} & \textbf{DeepEP} & \textbf{Ours}   \\
        
        \midrule
        \textbf{Intra-Node Swizzle} & \textcolor{yellow}{\ding{110}} & \textcolor{green}{\ding{51}} & \textcolor{green}{\ding{51}}  & \textcolor{yellow}{\ding{110}} & \textcolor{green}{\ding{51}} & \textcolor{green}{\ding{51}} & \textcolor{yellow}{\ding{110}} & \textcolor{green}{\ding{51}} \\

        \midrule
        \textbf{Inter-Node Swizzle} & \textcolor{yellow}{\ding{110}} & \textcolor{red}{\ding{55}} & \textcolor{red}{\ding{55}}  & \textcolor{yellow}{\ding{110}} & \textcolor{green}{\ding{51}} & \textcolor{green}{\ding{51}} & \textcolor{yellow}{\ding{110}} & \textcolor{green}{\ding{51}} \\

        \midrule
        \textbf{Inter-NUMA Swizzle} & \textcolor{yellow}{\ding{110}} & \textcolor{red}{\ding{55}} & \textcolor{red}{\ding{55}}  & \textcolor{red}{\ding{55}} & \textcolor{red}{\ding{55}} & \textcolor{red}{\ding{55}} & \textcolor{yellow}{\ding{110}} & \textcolor{green}{\ding{51}} \\

        \midrule
        \textbf{Copy Engine} & \textcolor{green}{\ding{51}} & \textcolor{green}{\ding{51}} & \textcolor{green}{\ding{51}}  & \textcolor{green}{\ding{51}} & \textcolor{green}{\ding{51}} & \textcolor{green}{\ding{51}} & \textcolor{red}{\ding{55}} & \textcolor{green}{\ding{51}} \\

        \midrule
        \textbf{High-BW Link} & \textcolor{green}{\ding{51}} & \textcolor{green}{\ding{51}} & \textcolor{green}{\ding{51}}  & \textcolor{green}{\ding{51}} & \textcolor{green}{\ding{51}} & \textcolor{green}{\ding{51}} & \textcolor{green}{\ding{51}} & \textcolor{green}{\ding{51}} \\

        \midrule
        \textbf{Network Comm.} & \textcolor{green}{\ding{51}} & \textcolor{green}{\ding{51}} & \textcolor{red}{\ding{55}}  & \textcolor{green}{\ding{51}} & \textcolor{green}{\ding{51}} & \textcolor{green}{\ding{51}} & \textcolor{green}{\ding{51}} & \textcolor{green}{\ding{51}} \\

        \midrule
        \textbf{PCIe Comm.} & \textcolor{green}{\ding{51}} & \textcolor{green}{\ding{51}} & \textcolor{red}{\ding{55}}  & \textcolor{red}{\ding{55}} & \textcolor{red}{\ding{55}} & \textcolor{green}{\ding{51}} & \textcolor{red}{\ding{55}} & \textcolor{green}{\ding{51}} \\

        \midrule
        \textbf{OpenSHMEM Support} & \textcolor{red}{\ding{55}} & \textcolor{red}{\ding{55}} & \textcolor{red}{\ding{55}}  & \textcolor{red}{\ding{55}} & \textcolor{red}{\ding{55}} & \textcolor{green}{\ding{51}} & \textcolor{green}{\ding{51}} & \textcolor{green}{\ding{51}} \\

        \midrule
        \textbf{Low-latency Protocol} & \textcolor{green}{\ding{51}} & \textcolor{red}{\ding{55}} & \textcolor{red}{\ding{55}}  & \textcolor{red}{\ding{55}} & \textcolor{red}{\ding{55}} & \textcolor{green}{\ding{51}} & \textcolor{red}{\ding{55}} & \textcolor{green}{\ding{51}} \\

        \midrule
        \textbf{Multimem Feature} & \textcolor{yellow}{\ding{110}} & \textcolor{red}{\ding{55}} & \textcolor{red}{\ding{55}}  & \textcolor{red}{\ding{55}} & \textcolor{red}{\ding{55}} & \textcolor{red}{\ding{55}} & \textcolor{red}{\ding{55}} & \textcolor{green}{\ding{51}} \\

        \midrule
        \textbf{Fusion} & \textcolor{red}{\ding{55}} & \textcolor{red}{\ding{55}} & \textcolor{red}{\ding{55}}  & \textcolor{yellow}{\ding{110}} & \textcolor{green}{\ding{51}} & \textcolor{green}{\ding{51}} & \textcolor{green}{\ding{51}} & \textcolor{green}{\ding{51}} \\

        \midrule
        \textbf{Code Generation} & \textcolor{red}{\ding{55}} & \textcolor{red}{\ding{55}} & \textcolor{red}{\ding{55}}  & \textcolor{green}{\ding{51}} & \textcolor{green}{\ding{51}} & \textcolor{red}{\ding{55}} & \textcolor{red}{\ding{55}} & \textcolor{green}{\ding{51}} \\

        \midrule
        \textbf{Nvidia/AMD} & \textcolor{green}{\ding{51}}/\textcolor{red}{\ding{55}} & \textcolor{green}{\ding{51}}/\textcolor{green}{\ding{51}} & \textcolor{green}{\ding{51}}/\textcolor{red}{\ding{55}}  & \textcolor{yellow}{\ding{110}}/\textcolor{red}{\ding{55}} & \textcolor{green}{\ding{51}}/\textcolor{red}{\ding{55}} & \textcolor{green}{\ding{51}}/\textcolor{red}{\ding{55}} & \textcolor{green}{\ding{51}}/\textcolor{red}{\ding{55}} & \textcolor{green}{\ding{51}}/\textcolor{green}{\ding{51}} \\

        \bottomrule
    \end{tabular}
\end{table}

Our Triton-distributed supports a wide range of optimization techniques. Although these optimizations are proposed or implemented originally in previous frameworks, Triton-distributed is the first to cover all these optimizations within one framework. We list 13 different optimizations that are required for different optimization purposes.

\begin{itemize}
    \item \textbf{Intra-Node Swizzle:} Swizzle is to change the order of communication operations and computation operations so that they can better overlap with other. Intra-node swizzle is to perform swizzling within a node.
    \item \textbf{Inter-Node Swizzle:} Inter-node swizzle is to perform swizzling across different nodes.
    \item \textbf{Inter-NUMA Swizzle:} For multi-socket systems, cross-NUMA communication performance is hard to optimize because of NUMA effect. Swizzling across different NUMAs could improve overlapping performance.
    \item \textbf{Copy Engine:} GPUs and NPUs employ dedicated memory copy units to perform communication. Using these units for communication is also important for overlapping within single node.
    \item \textbf{High-BW Link:} Utilizing high-bandwidth links such as NVLink and xGMI is critical for both intra-node and inter-node communication. Both copy engine and computing cores (e.g., SMs) can map communication operations to high-bandwidth links.
    \item \textbf{Network Communication:} Cross-node communication relies on network communication. Network communication optimization refers to map communication operations to network devices and schedule them to overlap with other operations.
    \item \textbf{PCIe Communication:} For accelerators that only support PCIe communication (e.g., L20), we need to schedule communications among PCIe links to avoid resource contention.
    \item \textbf{OpenSHMEM Support:} This optimization refers to using OpenSHMEM implementations (NVSHMEM or ROCSHMEM) to schedule communication operations.
    \item \textbf{Low-latency Protocol:} This optimization is to use low-latency protocol (proposed in NCCL) to achieve barrier-free communication.
    \item \textbf{Multimem Feature:} This optimization utilize hardware features to perform broadcast/reduction through dedicated instructions.
    \item \textbf{Fusion:} This optimization refers to fusing processing logics into communication, such as data casting, transposing, simple arithmetic operations, etc.
    \item \textbf{Code Generation:} This optimization refers to the ability to generate code just-in-time and to support tuning to further improve performance.
    \item \textbf{Nvidia/AMD:} This is used to represent hardware-specific optimizations. For Nvidia GPUs, optimizations include warpgroup MMA instructions, warp specialization, TMA instructions, etc. For AMD GPUs, optimizations include persistent kernel optimization and software pipelining.
\end{itemize}

In Table~\ref{table:optimization-support} we show the comparison of Triton-distributed to previous representative distributed overlapping frameworks. We use green check-mark (\textcolor{green}{\ding{51}}) to show that the framework supports the corresponding optimization; use yellow square (\textcolor{yellow}{\ding{110}}) to represent the corresponding optimization is potentially applicable in the framework, but not for sure; use red cross (\textcolor{red}{\ding{55}}) to represent that the optimization is not supported yet. Triton-distributed supported all the listed optimizations.


On NVIDIA H800 GPU clusters, we focus on three collective communication types: AllGather, ReduceScatter, and AllToAll. These three types are well supported in communication libraries such as NCCL~\cite{nccl}. However, these libraries perform synchronization before and after collective communication. As a result, overlapping is only available at the operator level (e.g., stream control and asynchronous wait). To achieve fine-grained overlapping, we need to break these collective communication operations into one-sided point-to-point communication operations and synchronize these operations with other computation operations through signals.

\subsection{Intra-node AllGather with Copy Engine}

For intra-node AllGather, we use copy engine for data transfer. Copy engine is a dedicated DMA (direct memory access) engine in GPU for data transfer between devices. Copy engine can be triggered using runtime interfaces such as \textit{CudaMemcpy} or \textit{CudaMemcpyAsync}.
For one-sided communication, depending on the direction of data transfer, there are two implementation variants: push mode and pull mode.
Using push mode, we can omit one synchronization operation but the data arrival order cannot be controlled; while using pull mode, we need an additional synchronization but the data arrival order can be controlled.

In Algorithm~\ref{algo:push-intra-ag}, we show the pseudo-code for push mode AllGather. \textit{remote\_ptr($T, r$)} is used to get symmetric remote buffer pointer of $T$ at rank $r$, which is then used to create a remote buffer through \textit{make\_buffer}.

\begin{algorithm}
\caption{One-sided Push-mode Intra-node AllGather}
\label{algo:push-intra-ag}

\begin{algorithmic}[1]
\State \textbf{Input:} Symmetric Buffer $T$, Signal $S$, Local Buffer $L$
\For{$r$ in range(WORLD\_SIZE)}
    \State remote\_buf = make\_buffer(remote\_ptr($T, r$) + RANK $\times L.size()$) 
    \State remote\_buf.copy\_($L, L.size()$) \textcolor{darkgreenRGB}{\textbf{// Memory Copy}}
    \State remote\_sig = remote\_ptr($S, r$) + RANK
    \State set\_signal(remote\_sig) \textcolor{darkgreenRGB}{\textbf{\ \ // Notify the consumer}}
\EndFor
\end{algorithmic}

\end{algorithm}

In Algorithm~\ref{algo:pull-intra-ag} we show pull mode one-sided AllGather. Compared to push mode, pull mode needs to perform local copy at first (line 3) and then uses a \textit{barrier\_all} to make the result of local copy visible to all the other ranks.
Then, other ranks are able to copy the data to their own buffer.

\begin{figure*}[t]
    \centering
\includegraphics[width=\textwidth]{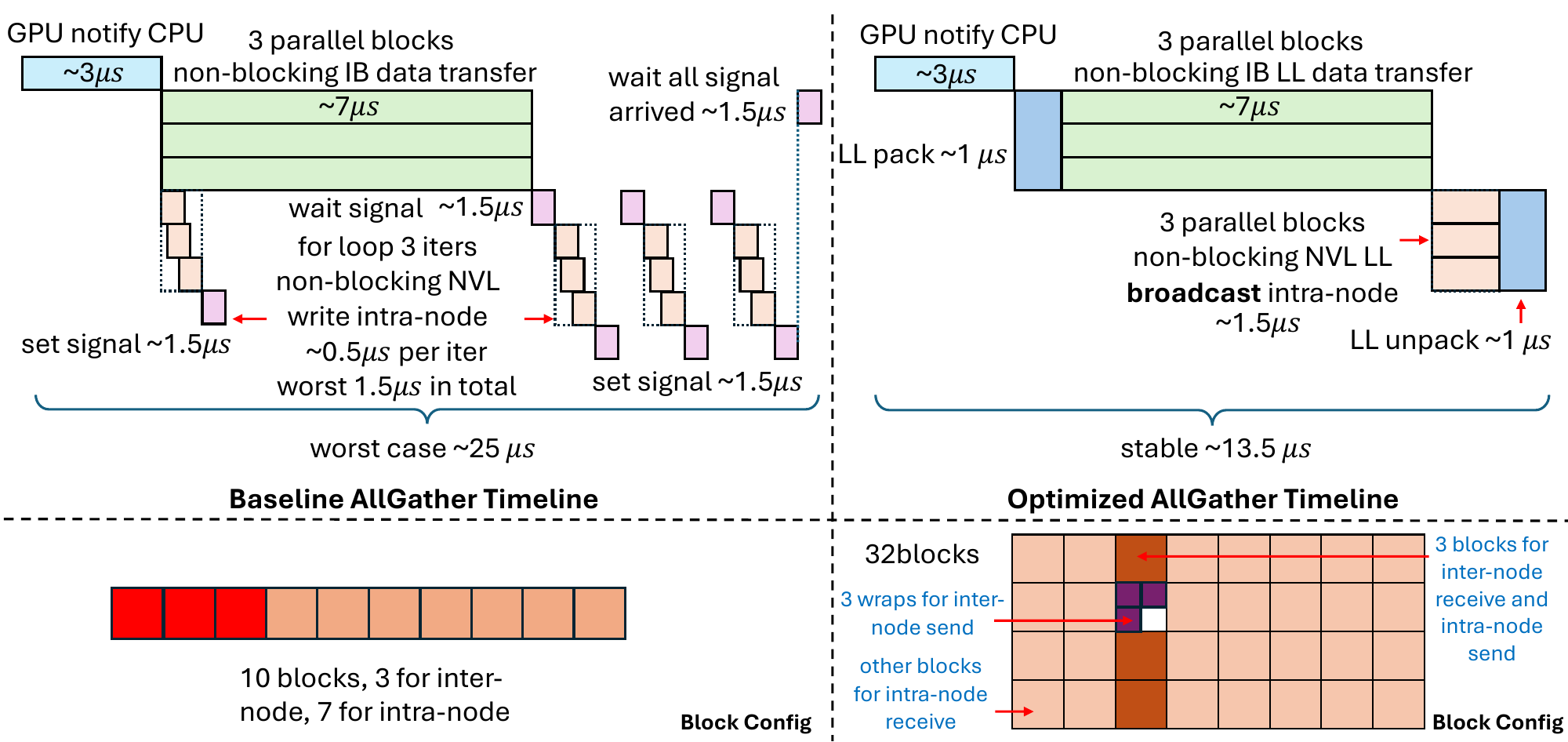}
    \caption{The Timeline of Baseline AllGather and Low-latency AllGather.}
    \label{fig:low-latency-allgather}
\end{figure*}

\begin{algorithm}
\caption{One-sided Pull-mode Intra-node AllGather}
\label{algo:pull-intra-ag}

\begin{algorithmic}[1]
\State \textbf{Input:} Symmetric Buffer $T$, Signal $S$, Local Buffer $L$
\State local\_t\_buf = make\_buffer($T$ + RANK $\times L.size()$) 
\State local\_t\_buf.copy\_($L, L.size()$)
\State set\_signal($S$ + RANK)
\State barrier\_all() \textcolor{darkgreenRGB}{\textbf{\ \ // Make the local copy visible to all the other ranks}}
\For{$r$ in range(WORLD\_SIZE)}
    \If{$r$ is not RANK}
        \State remote\_buf = make\_buffer(remote\_ptr($T, r$) + $r \times L.size()$)) 
        \State local\_t\_buf = make\_buffer($T$ + $r \times L.size()$) 
        \State local\_t\_buf.copy\_(remote\_buf, $L.size()$) 
        \State set\_signal($S + r$)
    \EndIf
\EndFor
\end{algorithmic}

\end{algorithm}

\subsection{Intra-node ReduceScatter with Copy Engine}

ReduceScatter is the reverse operation of AllGather. ReduceScatter can also be implemented in push or pull mode.
We only show push mode in Algorithm~\ref{algo:push-intra-rs}. The one-sided ReduceScatter is composed of two parallel parts. For the first part, local data shard is pushed to all the other ranks after the producer generates one tile of data; for the second part, local reduction is done and produces the final output. The two parts communicate with each other through signals.

\subsection{Inter-node AllGather with Low-latency Protocol and Multimem Feature} 

Inter-node AllGather requires overlapping the inter-node data transfer and the intra-node data transfer. For the code example in Figure~\ref{fig:ag-gemm-example-inter-node}, we have shown how to achieve such overlapping through assigning asyc-tasks to different threadblocks. However, this version of AllGather is only designed for high-bandwidth, not for low-latency.
For inference scenarios, low-latency AllGather is required for efficient parallel execution. The message size in AllGather is small, so processing delay and queuing delay are not critical. The main overhead of communication comes from propagation delay.

The AllGather implementation in Figure~\ref{fig:ag-gemm-example-inter-node} relies on loops to do data transfer, as a result, the data transfer operations are not launched at the same time. 
As shown on the left half of Figure~\ref{fig:low-latency-allgather}, using such an implementation, during the execution of AllGather, there is some skew among the data transmission operations sent to different ranks. Since the amount of transmitted data is very small, in the worst-case, the result of the skew is similar to sending data one by one, which leads to an extended overall delay. For example, the transmission via NVLink takes approximately 0.5 $\mu s$, but after the skew, it may take up to 1.5 $\mu s$ to transmit the data from the other 3 nodes at worst. In addition, each P2P data transfer requires a pair of signal operations (set signal and wait signal), causing additional overhead.

To address these issues and achieve low-latency AllGather, we propose to use non-OpenSHMEM primitives to achieve intra-node broadcast and use low-latency protocol (LL) for inter-node data transfer.
In detail, we use \textit{multimem\_st} primitive to do NVLink broadcast. The multimem instruction in Nvidia PTX instruction set is used to store the same data to all the other ranks within one node, which costs about 1.5$\mu s$. For inter-node communication, we use LL protocol, which relies on a hardware feature of Nvidia GPU that 8 bytes data store/load is atomic across ranks. Other low-latency protocol such as LL128 (which relies on the hardware feature of NVLink) can also be leveraged. Considering that we are targeting small message senarios, LL works perfect. To implement LL protocol, we store data and flags together into an 8 bytes data chunk and send the data directly to remote ranks, while the remote receivers use a spin-lock to check if the flag is the same as expectation to tell if the data has arrived.
LL protocol is fast but doubles the message size (due to flags in message), which is suitable for small message scenarios but not for large message size.

In Algorithm~\ref{algo:low-latency-allgather}, we show the pseudo-code of our low-latency AllGather. We use \textit{BLOCK\_ID} to denote the current threadblock index, use \textit{LOCAL\_WORLD\_SIZE} to denote the number of ranks in one node, use \textit{LOCAL\_RANK} to denote the rank index of the current device, use \textit{NODE\_ID} to denote the node index that the current device belongs to, use \textit{N\_NODES} to denote the number of nodes in total, and use \textit{WORLD\_SIZE} to denote the number of ranks from all the nodes in total.
We use \textit{recv\_LL\_pack} to perform LL receive operation without decoupling data from flags, while \textit{recv\_LL\_unpack} performs LL receive operation and separates the data from the flags.
In Algorithm~\ref{algo:low-latency-allgather}, we need totally \textit{WORLD\_SIZE} threadblocks and the block role configuration is shown in Figure~\ref{fig:low-latency-allgather}. One of the threadblock is responsible for inter-node data send and local data receive (line 11-18), three threadblocks are responsible for inter-node data receive and intra-node data send (line 6-9), while the other blocks are responsible for intra-rank receive (line 21-22).

The timeline of the low-latency AllGather is shown in the right part of Figure~\ref{fig:low-latency-allgather}. The estimated latency is 13.5 $\mu s$, which is better than the estimated latency of the baseline implementation (about 25 $\mu s$).

\begin{algorithm}
\caption{One-sided Push-mode Intra-node ReduceScatter}
\label{algo:push-intra-rs}

\begin{algorithmic}[1]
\State \textbf{Input:} Symmetric Buffer $T$, Signal from producer $P$, Signal for Reduction $S$, Local Buffer $L$, Reduce Output buffer $R$
\State \textcolor{darkgreenRGB}{\textbf{// In Stream 1}}
\For{$r$ in range(WORLD\_SIZE)}
    \State wait\_signal($P$ + RANK) \textcolor{darkgreenRGB}{\textbf{\ \ // Wait for the producer to generate one tile of data}}
    \State remote\_buf = make\_buffer(remote\_ptr($T, r$) + RANK $\times R.size()$) 
    \State remote\_buf.copy\_($L$ + RANK $\times R.size(), R.size()$)
    \State remote\_sig = remote\_ptr($S, r$) + RANK
    \State set\_signal(remote\_sig)
\EndFor
\State \textcolor{darkgreenRGB}{\textbf{// In Stream 2}}
\For{$r$ in range(WORLD\_SIZE)}
    \State wait\_signal($S + r$)
    \State tiled\_buf = make\_buffer($T + r \times R.size()$)
    \State $R = R + $tiled\_buf
\EndFor
\end{algorithmic}

\end{algorithm}

\begin{algorithm}
\caption{One-sided Low-latency Cross-Node AllGather}
\label{algo:low-latency-allgather}

\begin{algorithmic}[1]
\State \textbf{Input:} Symmetric Buffer $T$, Low-latency Buffer $L$, Signal $P$, Bytes per Rank $bytes$
\State peer\_node\_id = BLOCK\_ID / LOCAL\_WORLD\_SIZE
\State peer\_local\_rank = BLOCK\_ID \% LOCAL\_WORLD\_SIZE
\If{peer\_local\_rank == LOCAL\_RANK}
    \If{NODE\_ID != peer\_node\_id}
        \State seg = peer\_node\_id $\times$ LOCAL\_WORLD\_SIZE + LOCAL\_RANK
        \State recv\_LL\_pack($L$ + seg $\times bytes \times 2$, $L$ + seg $\times bytes \times 2$)
        \State multimem\_st($L$ + seg $\times bytes \times 2$, $L$ + seg $\times bytes \times 2$)
        \State recv\_LL\_unpack($T$ + seg $\times bytes$, $L$ + seg $\times bytes \times 2$)
    \Else
        \State LL\_pack($L$ + RANK $\times bytes \times 2$, $T$ + RANK $\times bytes$)
        \State \_\_syncthreads()
        \If{WARP\_ID < N\_NODES\ and\ WARP\_ID != NODE\_ID}
            \State peer = WARP\_ID $\times$ LOCAL\_WORLD\_SIZE + LOCAL\_RANK
            \State putmem\_nbi\_warp($L$ + RANK $\times bytes \times 2$, $L$ + RANK $\times bytes \times 2$, peer)
        \EndIf
        \State seg = peer\_node\_id $\times$ LOCAL\_WORLD\_SIZE + LOCAL\_RANK
        \State multimem\_st($L$ + seg $\times bytes \times 2$, $L$ + seg $\times bytes \times 2$)
    \EndIf
\Else
    \State seg = peer\_node\_id $\times$ LOCAL\_WORLD\_SIZE + peer\_local\_rank
    \State recv\_LL\_unpack($T$ + seg $\times bytes$, $L$ + seg $\times bytes \times 2$)
\EndIf
\end{algorithmic}

\end{algorithm}

\subsection{Inter-node ReduceScatter with Heterogeneous Communication}

\begin{algorithm}
\caption{One-sided Push-mode Cross-node ReduceScatter}
\label{algo:push-inter-rs}

\begin{algorithmic}[1]

\State \textbf{Input:} Local Buffer $L[M, N]$, Signal from producer $P$, Signal for inter node communication $S$, Reduce Output buffer $R$

\State \textbf{Input:} partial\_rs\_buf = Symmetric Buffer(N\_NODES, M\_PER\_RANK, N)
\State \textbf{Input:} scatter\_buf = Symmetric Buffer(LOCAL\_WORLD\_SIZE, M\_PER\_RANK, N)
\State M\_PER\_RANK = M / WORLD\_SIZE
\State M\_PER\_NODE = M / N\_NODES
\For{$n$ in range(N\_NODES)}
    \State \textcolor{darkgreenRGB}{\textbf{// In Stream 0: intra node scatter}}
    \For{$r$ in range($LOCAL\_WORLD\_SIZE$)}
        \State gr = r + $NODE\_ID$ * $LOCAL\_WORLD\_SIZE$
        \State wait\_signal($P$ + gr)
        \State remote\_buf = make\_buffer(remote\_ptr(scatter\_buf[$LOCAL\_RANK$], gr))
        \State remote\_buf.copy\_($L$ + gr $\times$ M\_PER\_RANK $\times$ N, M\_PER\_RANK $\times$ N)
    \EndFor
    \State barrier\_all\_intra\_node(stream0)

    \State \textcolor{darkgreenRGB}{\textbf{// In Stream 1: local reduction and inter node P2P}}
    \State stream1.wait(stream0)
    \State rs\_cur\_node = Reduce(scatter\_buf, dim=0)
    \State target\_rank = $LOCAL\_RANK$ + n $\times LOCAL\_WORLD\_SIZE$
    \State P2P\_send(rs\_cur\_node, remote\_ptr(partial\_rs\_buf[$NODE\_ID$], target\_rank))
\EndFor
\State barrier\_all()
\State R = Reduce(partial\_rs\_buf, dim=0)

\end{algorithmic}

\end{algorithm}

The inter-node ReduceScatter can be decomposed into three stages: namely intra-node scatter, local reduction, and inter-node P2P communication. Since the local reduction operation requires SM resources, we aim to maximize bandwidth while minimizing resource usage to ensure that the computation (e.g., GEMM) performance is affected as little as possible.

in Algorithm ~\ref{algo:push-inter-rs}, we show the pseudo-code of our inter-node ReduceScatter. We use P2P\_send to denote inter-node P2P communication. To optimize resource usage and enhance communication efficiency, we employ an overlapping strategy for intra-node and inter-node communications. We schedule the intra-node scatter on one stream, while the local reduction and P2P communication are assigned to another stream. The scatter operation is completed by the copy engine and does not require SM; P2P communication only requires 1 SM, and the number of SMs for local reduction is the minimum required value calculated based on hardware specifications.

Taking 8$\times$H800 as an example (around 170 GB/s NVLink maximum bandwidth), each GPU is connected to a CX7 InfiniBand 400 Gb/s RDMA network card (around 45 GB/s maximum bandwidth), the communication volume is $B = \frac{M\_PER\_RANK \times N\times sizeof(dtype)}{10^9}$ GB. The time for P2P communication is $\frac{B}{45}$, and the time for scatter is $\frac{(8-1)\times B}{170}$. So the left time for reduction is $\frac{(8-1)\times B}{170} - \frac{B}{45}$. It can be inferred that if the bandwidth of the local reduction operator exceeds 470 GB/s, perfect overlap can be achieved. Therefore, on H800 GPU, it is sufficient to assign only a small number of SMs (no more than 15) for local reduction.

\subsection{Optimized Communication Kernels on AMD GPUs and more Platforms}

On AMD MI308X GPU cluster, we focus on AllGather and ReduceScatter collective communications. Similarly, RCCL~\cite{rccl} is the typical communication library used on AMD GPUs to perform these collective operations, and often coupled with synchronizations before and after execution. Thus the overlapping is limited within scope of operator-level. We employ similar signal-based approach to synchronize computation and communication to achieve fine-grained overlapping.

\textbf{Intra-node AllGather with Copy Engine}

For intra-node AllGather on AMD, we still utilize copy engine for data transfer between different GPUs. Runtime APIs such as \textit{hipMemcpyAsync} is used to control copy engine on AMD GPU, and options like \textit{hipMemcpyDeviceToDeviceNoCU} can be used to specify not use GPU compute units (CU) to avoid affecting computations happen at the same time.

Unlike Nvidia H800 cluster that GPU connected through NVLink system, on AMD MI308X cluster each GPU connects with others via bidirectional link. Therefore, to maximize the bandwidth, the data transfer needs to be launched on multiple streams simultaneously.
After the data transfer, corresponding signals need to be set to notify the consumers. Typically these can be achieved using driver APIs such as \textit{cuStreamWriteValue} or \textit{hipStreamWriteValue} on Nvidia and AMD GPUs respectively. However, these driver APIs seem to interfere with computation kernels and cause considerable delay on AMD GPUs. Thus we have to use other approach to workaround, here we launched another data transfer solely for signals.

Besides the push or pull mode variants for one-sided communication, communication tile size also plays an important factor in optimization.
In AllGather, communication tiling is decoupled from computation tiling to avoid any interference. Tuning communication tiling independently allows us to find a best trade off between overlapping opportunity and communication efficiency, minimizing effective communication time.
Moreover the communication tile order also matters, it needs to align with corresponding consumer computation to minimizing overall delay. In other words, it requires the computation tile coordinate swizzling to align with the arriving order of the communication tile.

\textbf{Intra-node ReduceScatter with Fusion}

For ReduceScatter, due to the previous mentioned interfere issue, we fused scatter part with producer kernel to store the output data directly to other ranks once the tile of data gets ready and avoid the need for any driver API. Then the reduction part is performed follow by \textit{barrier\_all} and produce the final output.

Here the producer kernel has to be modified to decouple the tile size with communication tile, and add the communication tile size as another tuning factor in autotune. Again the tile coordinate swizzling needs to align with communication tile order to utilize all the bidirectional links so as to maximize the bandwidth.

\textbf{Optimizations on More Platforms}
Although we only show optimized kernels for GPUs currently, we can support other platforms similarly. To achieve this, we require the target hardware to support the three core concepts in our programming model.
First, the hardware should support symmetric memory allocation and access, which is expected to conform to the OpenSHMEM standard.
Second, the hardware should support signal exchange, including signal setting, increasing, checking, and spin-locks.
Third, the hardware should support async-tasks, allowing us to map specialized tasks spatially to different hardware units.
With the support of the three core concepts, we can port our OpenSHMEM and non-OpenSHMEM primitives to the hardware platform accordingly.

\subsection{Overlapping Computation with Swizzling Optimization}

\begin{figure}[t]
    \centering
\includegraphics[width=0.9\textwidth]{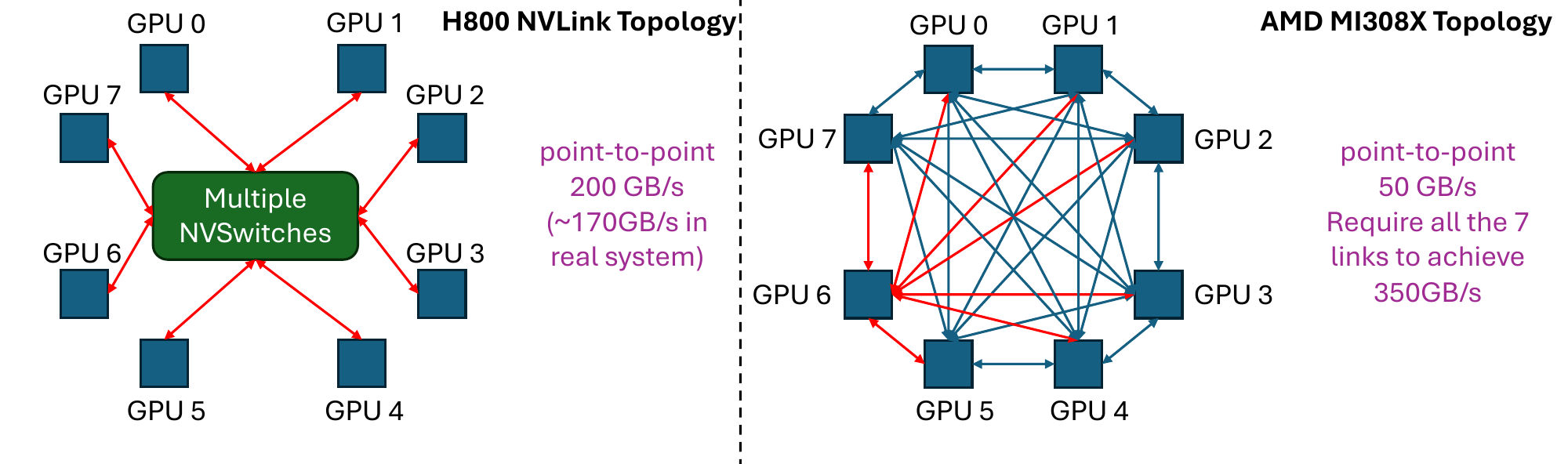}
    \caption{Topology of Nvidia GPUs and AMD GPUs}
    \label{fig:gpu-topo}
\end{figure}

\begin{figure}[t]
    \centering
\includegraphics[width=0.8\textwidth]{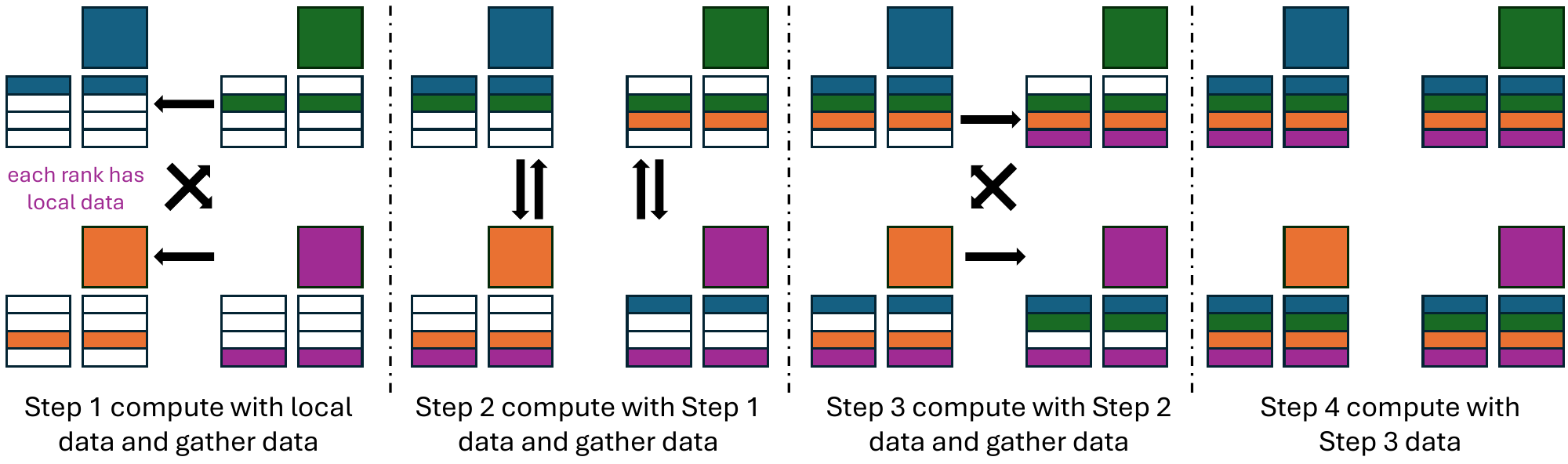}
    \caption{Swizzle Example for Intra-node AllGather GEMM on Nvidia GPUs. (Assume 4 ranks.)}
    \label{fig:ag-gemm-swizzle}
\end{figure}

\begin{figure}[t]
    \centering
\includegraphics[width=0.9\textwidth]{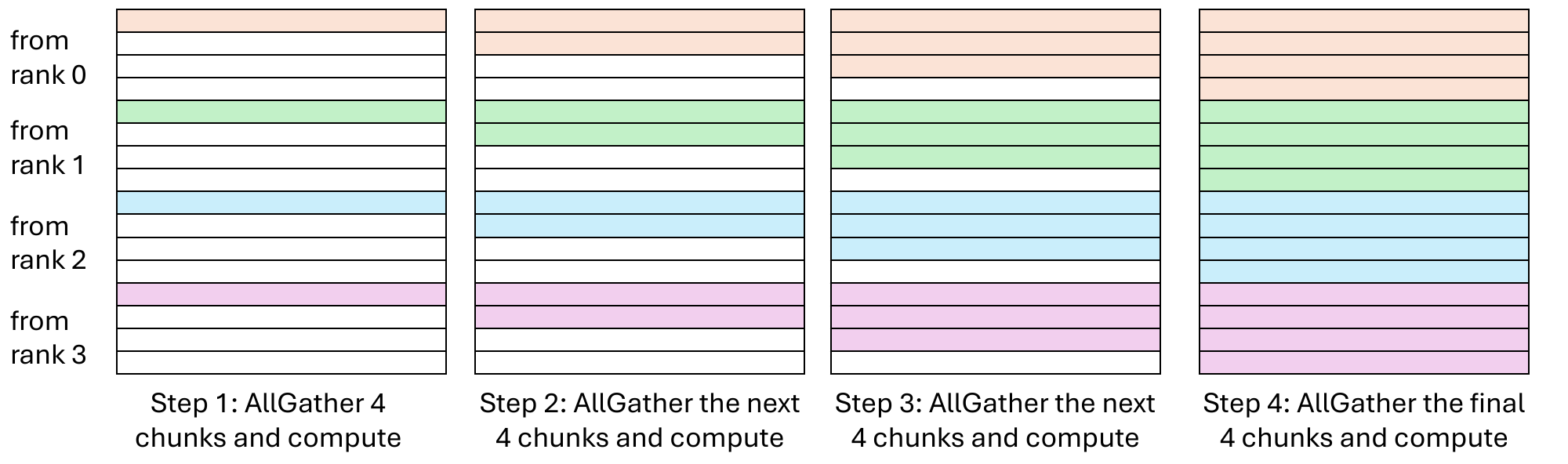}
    \caption{Swizzle Example for Intra-node AllGather GEMM on AMD GPUs. (Assume 4 ranks.)}
    \label{fig:ag-gemm-swizzle-amd}
\end{figure}

\begin{figure}[t]
    \centering
\includegraphics[width=0.9\textwidth]{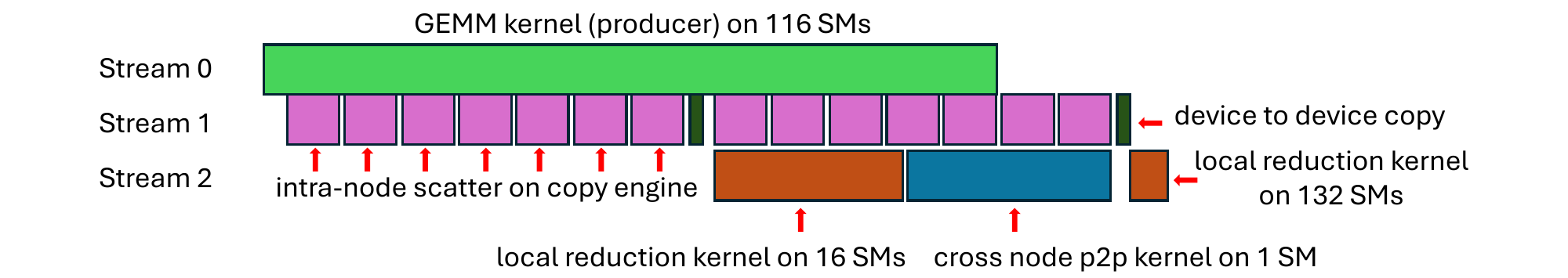}
    \caption{The Timeline of Inter-node GEMM ReduceScatter.}
    \label{fig:gemm-rs-timeline}
\end{figure}

\begin{figure*}[tb]
    \centering
\includegraphics[width=0.7\textwidth]{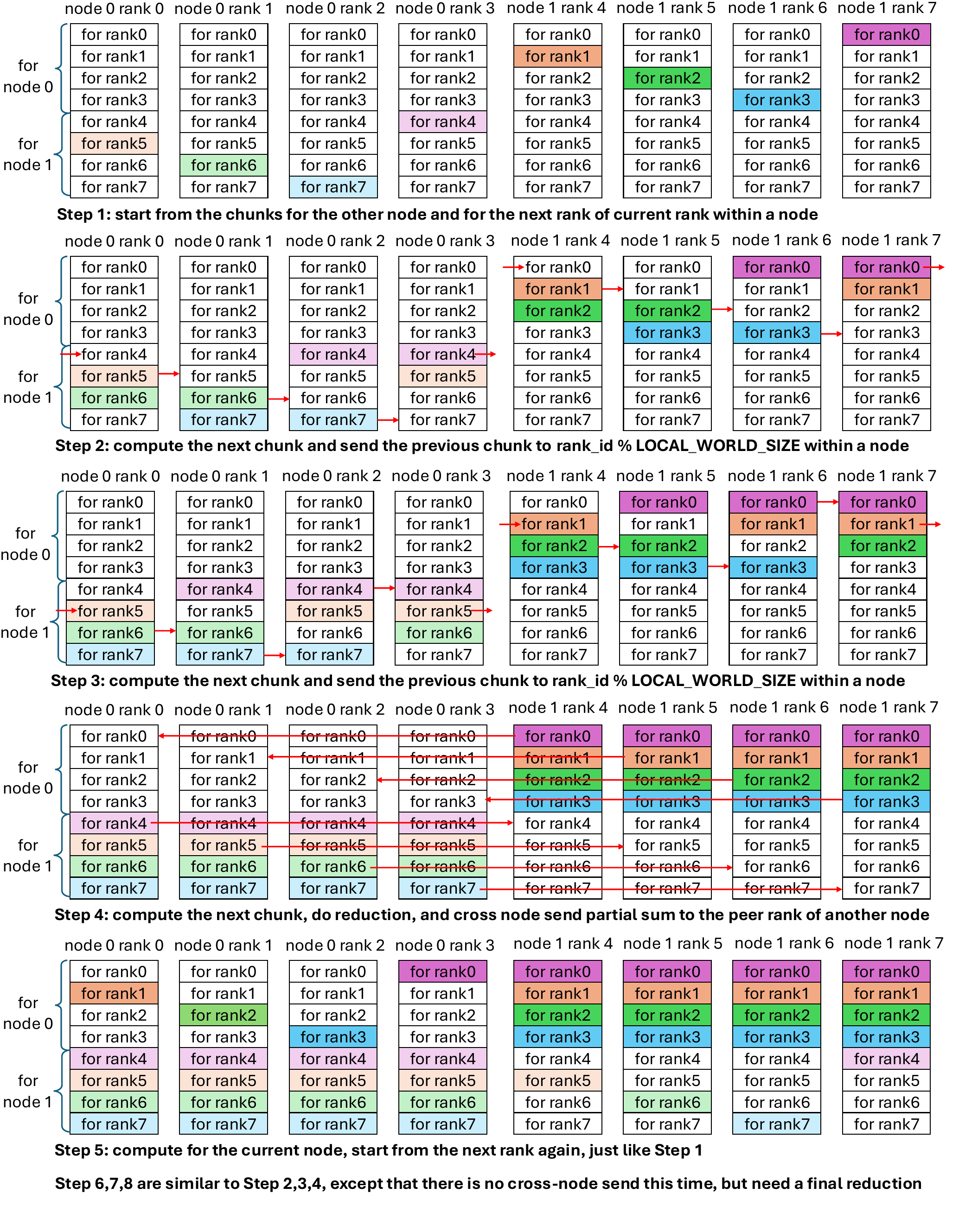}
    \caption{Swizzle Example for Inter-node GEMM ReduceScatter. (Assume 2 nodes, each with 4 ranks.)}
    \label{fig:gemm-rs-swizzle}
\end{figure*}

\begin{table}[h]
    \centering
    \footnotesize
    \caption{List of Our Optimized Kernels}
    \label{table:supported-kernels}
    \begin{tabular}{c|c|c}
        \toprule
        \textbf{Name} & \textbf{Explanation} & \textbf{Tested Hardware Cluster} \\
        \midrule
        AG+GEMM-intra & Intra-node AllGather GEMM Overlapping & 8 H800 and MI308X GPUs.\\
        \midrule
        GEMM+RS-intra & Intra-node GEMM ReduceScatter Overlapping & 8 H800 and MI308X GPUs.\\
        \midrule
        AG+MoE-intra & Intra-node AllGather MoE GroupGEMM Overlapping & 8 H800 GPUs\\
        \midrule
        MoE+RS-intra & Intra-node MoE GroupGEMM ReduceScatter Overlapping & 8 H800 GPUs\\
        \midrule
        FlashDecode+AG-intra & Intra-node Flash Decode AllGather and Combine & 8 H800 GPUs\\
        \midrule
        AllToAll-intra & Intra-node Low-latency AllToAll & 8 H800 GPUs\\
        \midrule
        AG+GEMM-inter & Inter-node AllGather GEMM Overlapping & 16 H800 GPUs\\
        \midrule
        GEMM+RS-inter & Inter-node GEMM ReduceScatter Overlapping & 16 H800 GPUs\\
        \midrule
        AG+MoE-inter & Inter-node AllGather MoE GroupGEMM Overlapping & 16 H800 GPUs\\
        \midrule
        MoE+RS-inter & Inter-node MoE GroupGEMM ReduceScatter & 16 H800 GPUs\\
        \midrule
        FlashDecode+AG-inter & Inter-node Flash Decode AllGather and Combine & 16 and 32 H800 GPUs\\
        \midrule
        AllToAll-inter & Inter-node Low-latency AllToAll & 16, 32, and 64 H800 GPUs\\

        \bottomrule
    \end{tabular}
\end{table}


So far, we have mainly discussed collective communication parts of optimized overlapping kernels. In this section, we introduce the optimizations for overlapping the computation part.
Efficient GPU kernels, either Nvidia or AMD GPU, both rely on tiling to exploit parallelism and locality. And there is a tile mapping logic, from a thread block index to a tile coordinate. 
By controlling the order of tiles, we can both improve cache utilization (e.g., L2 cache) and communication efficiency (by reducing the critical path).
The optimization that controls the order of tiles is called swizzling.

The design of tile swizzle requires the tile swizzle order in computation kernels align with the communication tiles for two purposes: 1) avoid potential memory contention and 2) minimize tile data transfer time~\cite{flux}.
Moreover, because of the complexity of topology connections between GPUs in one node or multi-nodes, as well as different vendors, tile swizzling in computation kernels has to consider all these variants that reflects in collective communication to maximize overlapping and minimize overall latency.

For Nvidia GPUs and AMD GPUs, they use different topology for interconnection. In Figure~\ref{fig:gpu-topo}, we show the difference between them.
Nvidia H800 GPUs use NVSwitches to connect the 8 GPUs within one node. Each pair of GPUs can communicate with each other with a maximum of 200 GB/s uni-direction bandwidth. But AMD MI308X GPUs use a full-mesh topology, where each GPU connects to the other 7 GPUs with 7 different links. Each link has a maximum of 50 GB/s uni-direction bandwidth. The aggregated bandwidth for a single GPU is 350 GB/s.

Different operators require different swizzle methods to fully utilize the interconnection bandwidth. Without loss of generality, we show the swizzle methods for AllGather GEMM and GEMM ReduceScatter. The swizzle methods for other operators can be designed similarly.
For AllGather GEMM on Nvidia GPUs, we show the swizzle in Figure~\ref{fig:ag-gemm-swizzle}. We use a small number of ranks (4 ranks) for simplicity. At the beginning of AllGather GEMM, each rank occupies a chunk of data (local data). For the first step, each rank uses its own local data to compute part of outputs, and at the same time, each rank P2P gathers the next chunk data from another rank. So the GEMM kernel of each rank starts from a unique tile index, which is relevant to its own rank index. For the second step, each rank uses the data gathered from the previous step to compute part of outputs, and meanwhile, P2P gathers the next chunk of data from another rank. As the bandwidth of each link of NVLink can reach the peak bandwidth, each rank only gathers data from one another rank at a time. 

For AllGather GEMM on AMD GPUs, the swizzle design is different. If we only gather the next chunk of data from another one GPU at a time, the maximal available bandwidth is only 50 GB/s. To fully utilize the bandwidth, each rank should gather data chunks from all the other ranks at each step. In Figure~\ref{fig:ag-gemm-swizzle-amd} we show the sizzle method for AMD GPUs. For each chunk of data in each rank, we further tile them into sub-chunks. For each step, each rank gathers all the next sub-chunks from all the other ranks. Figure~\ref{fig:ag-gemm-swizzle-amd} shows the swizzle view from a single rank. For Step 1, it gathers the first sub-chunk from all the other ranks and complete the GEMM computation using these sub-chunks. For the second step, it gathers the next set of sub-chunks. The remaining steps follow a similar pattern.

Besides the design for intra-node topology, we also provide swizzling design for inter-node overlapping. We use GEMM ReduceScatter for illustration as this example is the most representative one.
For simplicity, we use 2 nodes with totally 8 ranks (4 ranks per node).
The swizzling steps are shown in Figure~\ref{fig:gemm-rs-swizzle}.
We decompose the GEMM ReduceScatter into three stages as explained previously. For intra-node scatter stage, each rank performs 7 remote data transfers and one local data copy (for its own rank index). The intra-scatter stage is repeated 2 times for each rank.
To overlap the most of communication, we should arrange the local copy step of intra-node scatter to the tailing position of the stage. So each rank starts its own computation from the next chunk of data relative to its rank index, rather than from its own rank index.
Similarly, to overlap inter-node P2P data transfer, each node should start its computation from the data chunks needed by the other node.
As a result, the Step 1 in Figure~\ref{fig:gemm-rs-swizzle} uses a shift for the starting position for each tile of computation and each chunk of data.
In detail, rank 0 starts its GEMM for the data required by rank 5, rank 1 starts computation for rank 6, and so forth.
After each step of computation, each rank sends its partial sum to the next rank within the same node for local reduction.
After all the ranks within one node get all the chunks of data, each rank P2P sends the partial sum to the peer rank in another node (Step 5). After this, all the ranks within the same node repeat the Steps 2-4 to produce the final output.

\subsection{Code Generation Optimization with Auto-Tuning and Resource Partition}

Besides swizzling optimization, we also leverage two important optimizations for better performance.
The first is tiling factor tuning, and the second is resource partition.


For tiling factor tuning, we develop an autotuner tailored for distributed kernels. The autotuners used in previous compilers~\cite{tvm, triton} iteratively launch a single kernel with different tiling configurations on a single device to discover the best one. However, unlike tuning on a single device, distributed kernel tuning involves the communication and synchronization between different kernels across multiple devices. This requires us to consider both the synchronization needs of the kernel launch itself and the synchronization of the tuning results. For example, since we focus on overlapping kernels, we need to reset all the signals every time we profile the generated code, and we cannot just repeatedly execute the target kernel like the previous autotuner, as this would disrupt the synchronization conditions.
To this end, our autotuner can accept a target function that wraps the overlapping kernels (encompassing communication, computation, and host-side launch logic), including the kernels requiring tuning. The target function is executed iteratively as a whole. In each iteration, the kernel that needs tuning within the target function is executed only once. The profiling and enumeration of tuning configurations are progressively carried out as the target function iterates. Finally, upon completion of a kernel's tuning, a global synchronization is performed to aggregate the tuning information from different devices, thereby selecting a globally unified best configuration.

Resource partition is to spatially map the computation and communication to different processing units (e.g. SMs).
For intra-node overlapping kernels, the communication can be mapped to copy engine, and the computation can fully utilize all the compute cores (SMs or CUs).
For inter-node overlapping kernels, communication also requires computing cores to complete remote memory access. And the insight in resource partition is to make sure all the async-tasks can overlap with each other and complete at the same time (i.e. avoid long tails). Take inter-node GEMM ReduceScatter introduced previously as example, the timeline of our overlapping kernels is shown in Figure~\ref{fig:gemm-rs-timeline}, on H800 GPUs, the GEMM kernel uses 116 SMs, the intra-node scatter is mapped to copy engine, the cross node P2P kernel uses 1 SM, the first local reduction kernel uses 16 SMs, and the second local reduction kernel uses 132 SMs. With this configuration, perfect overlapping can be achieved.

\section{Experiments}

In this section, we demonstrate the performance of our optimized kernels. We list the supported kernels in Table~\ref{table:supported-kernels}. We use both Nvidia H800 GPUs and AMD MI308X GPUs to test the performance. The scale of our cluster ranges from 8 GPUs to 64 GPUs.
For each case, we list the problem shapes and baseline accordingly.

\subsection{Intra-node Kernel Performance on Nvidia GPUs}

\textbf{AllGather GEMM}

For AG+GEMM-intra, we compare with PyTorch+NCCL (PyTorch uses cuBLAS~\cite{cublas} for GEMM) and FLUX~\cite{flux} (FLUX uses CUTLASS~\cite{cutlass} for GEMM). For GEMM performance, Triton's generated code can achieve roughly ~95\% the performance of cuBLAS and CUTLASS. Although our GEMM performance is not the best, we still achieve better AG+GEMM performance due to enhanced overlapping.
On average, we achieve $1.42\times$ speedup to PyTorch+NCCL and $1.09\times$ speedup to FLUX.

\begin{figure}[t]
    \centering
\includegraphics[width=0.8\textwidth]{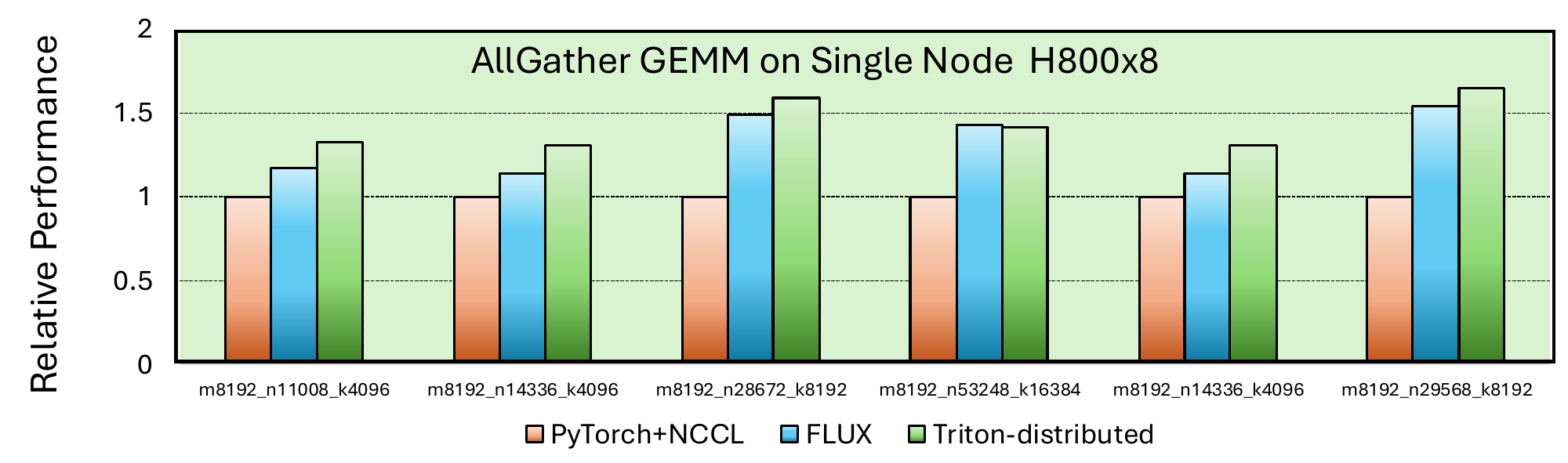}
    \caption{Performance of Intra-node AllGather GEMM on 8 H800 GPUs.}
    \label{fig:ag-gemm-intra-node-perf}
\end{figure}

\begin{figure}[t]
    \centering
\includegraphics[width=0.8\textwidth]{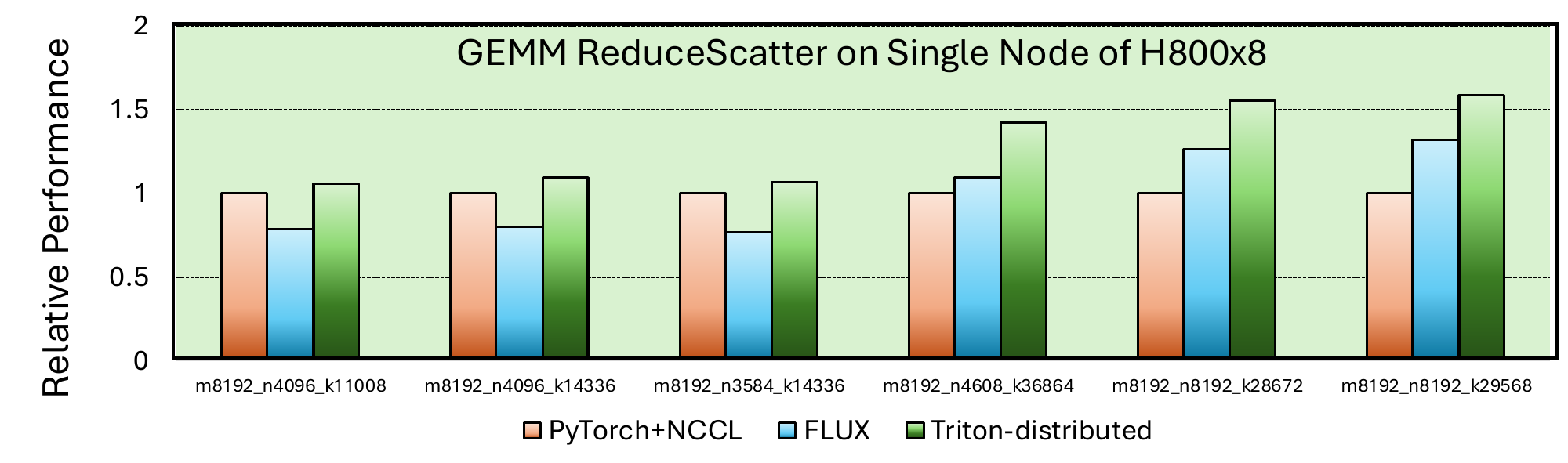}
    \caption{Performance of Intra-node GEMM ReduceScatter on 8 H800 GPUs.}
    \label{fig:gemm-rs-intra-node-perf}
\end{figure}

\begin{figure}[t]
    \centering
\includegraphics[width=0.8\textwidth]{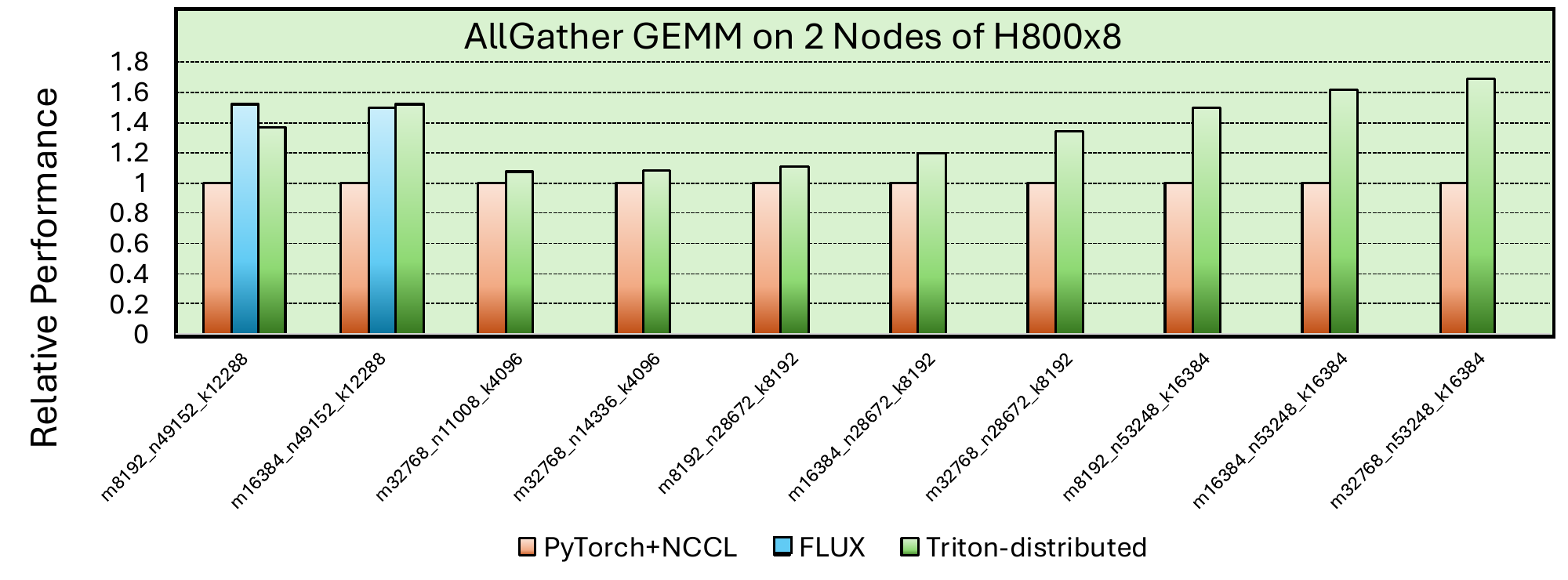}
    \caption{Performance of Inter-node AllGather GEMM on 16 H800 GPUs.}
    \label{fig:ag-gemm-inter-node-perf}
\end{figure}

\begin{figure}[t]
    \centering
\includegraphics[width=0.8\textwidth]{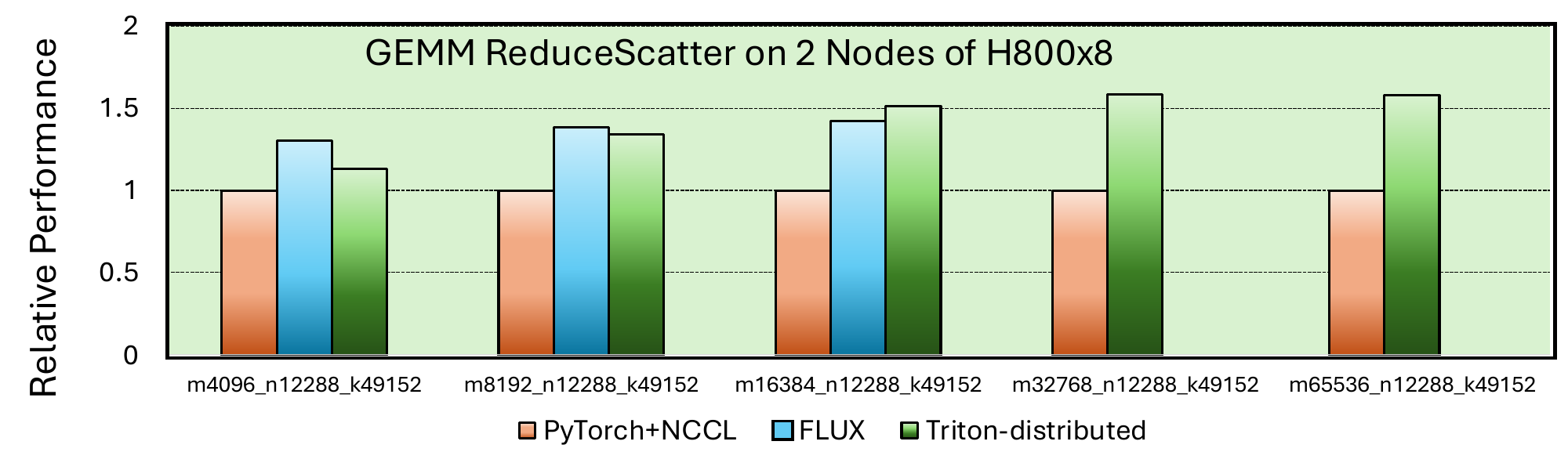}
    \caption{Performance of Inter-node GEMM ReduceScatter on 16 H800 GPUs.}
    \label{fig:gemm-rs-inter-node-perf}
\end{figure}

\begin{figure}[htb]
    \centering
\includegraphics[width=0.8\textwidth]{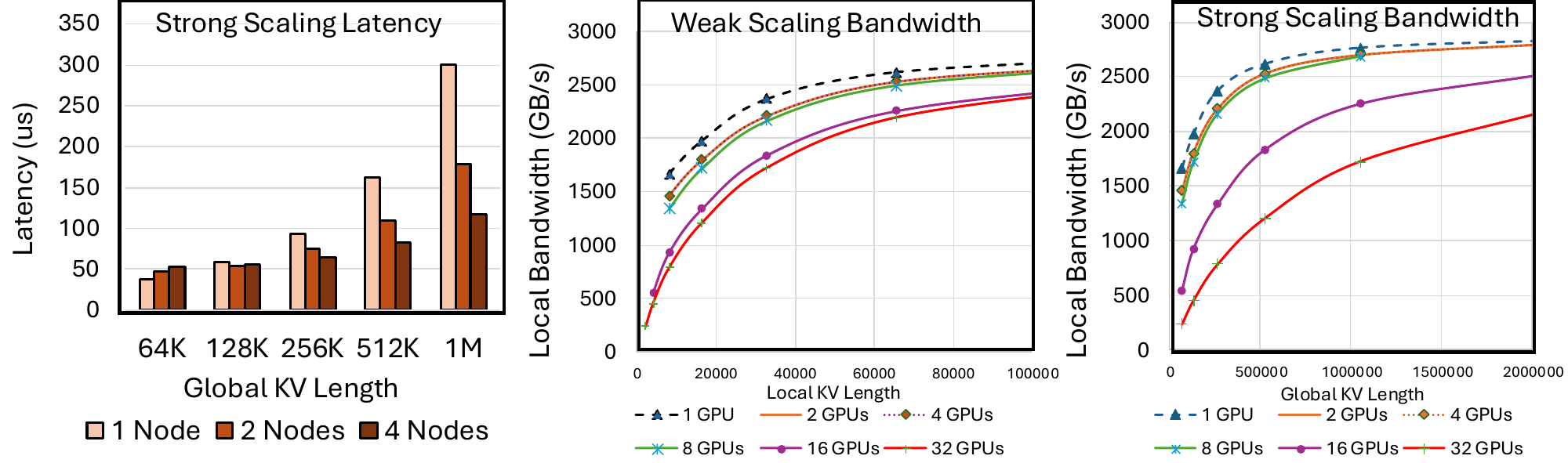}
    \caption{Performance of Distributed Flash Decoding.}
    \label{fig:flash-decode-perf}
\end{figure}

\begin{figure}[htb]
    \centering
\includegraphics[width=0.8\textwidth]{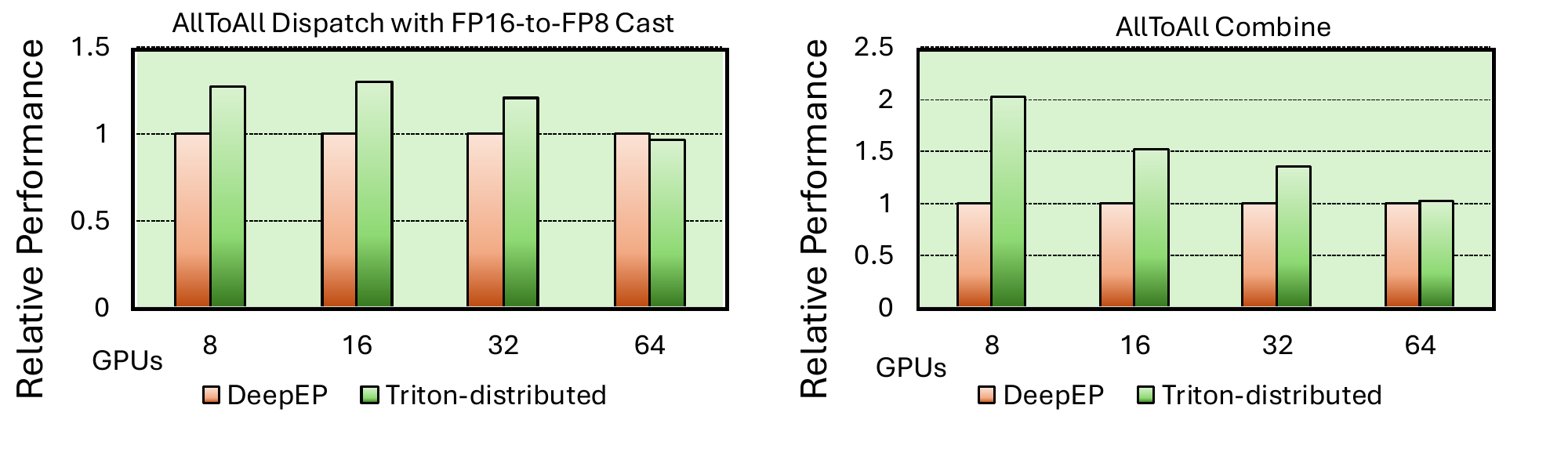}
    \caption{Performance of AllToAll.}
    \label{fig:all2all-perf}
\end{figure}

\begin{figure}[htb]
    \centering
\includegraphics[width=0.8\textwidth]{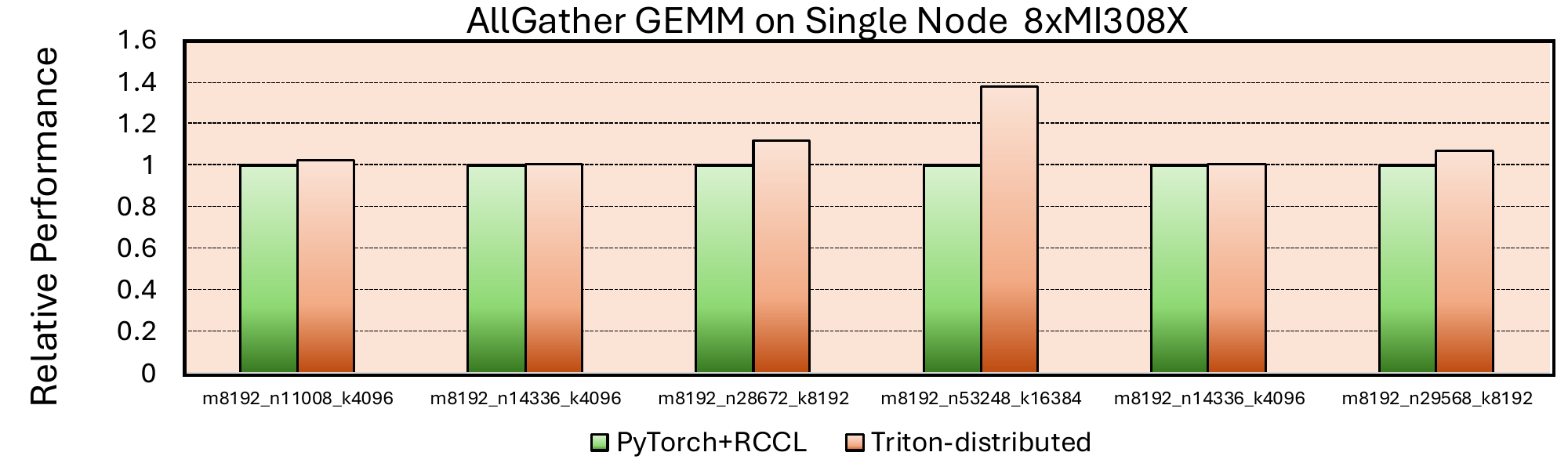}
    \caption{Performance of Intra-node AllGather GEMM on AMD GPUs.}
    \label{fig:amd-ag-gemm-perf}
\end{figure}

\begin{figure}[htb]
    \centering
\includegraphics[width=0.8\textwidth]{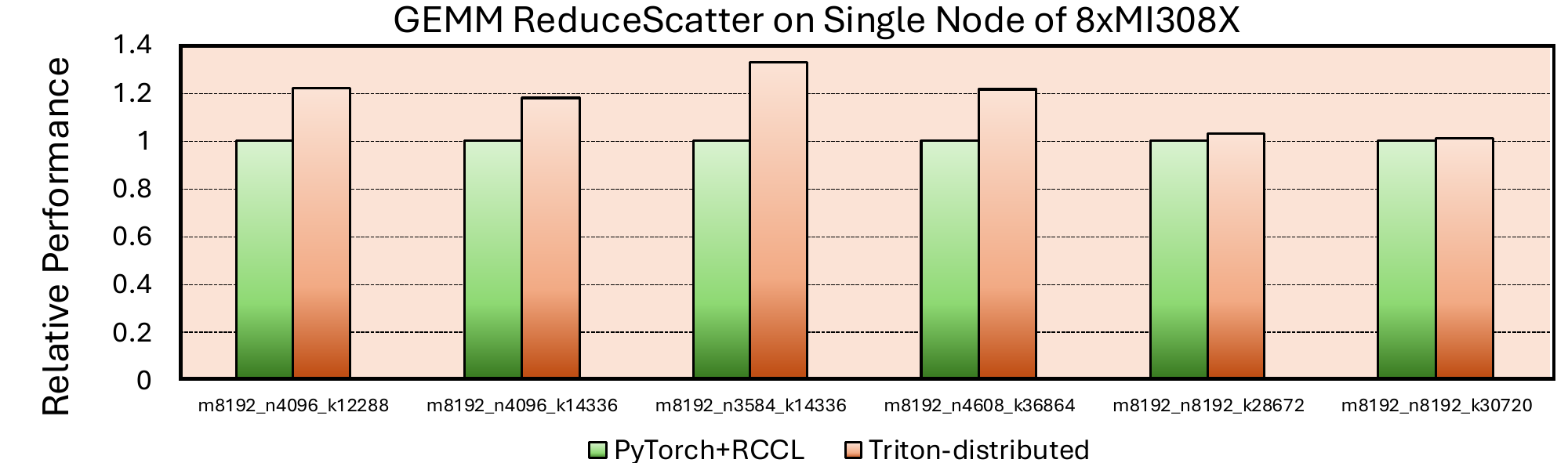}
    \caption{Performance of Intra-node GEMM ReduceScatter on AMD GPUs.}
    \label{fig:amd-gemm-rs-perf}
\end{figure}

\begin{figure}[htb]
    \centering
\includegraphics[width=1.0\textwidth]{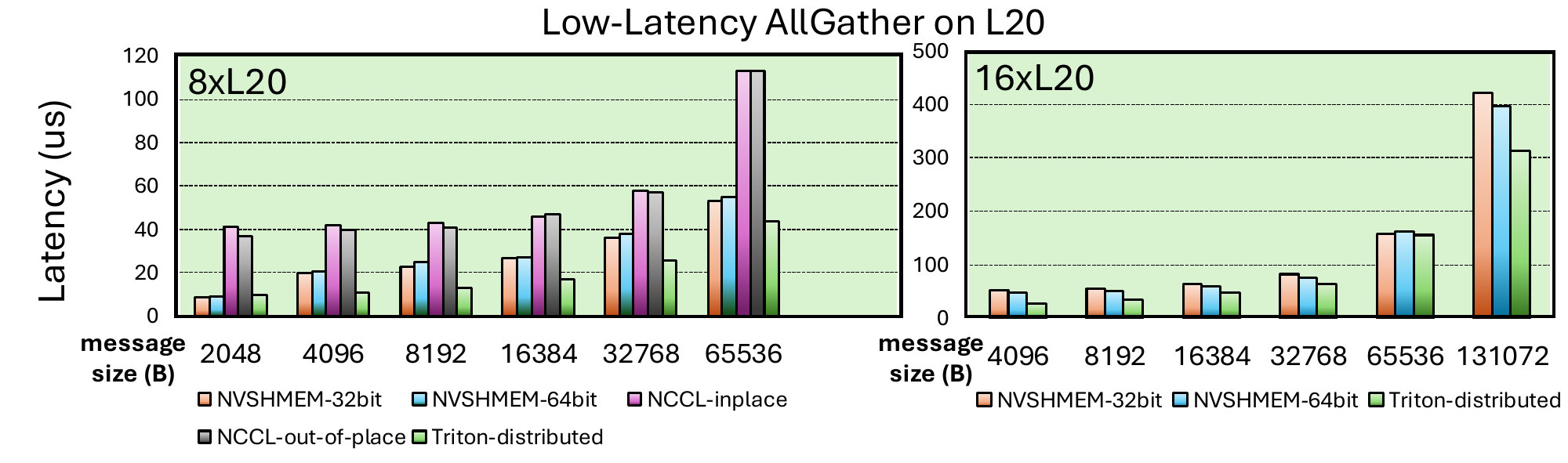}
    \caption{Performance of Low-latency AllGather on L20 GPUs.}
    \label{fig:low-latency-ag-l20}
\end{figure}

\textbf{GEMM ReduceScatter}

For GEMM+RS-intra, we also compare to PyTorch+NCCL and FLUX. Our overlapping design is different from FLUX. FLUX fuses the scatter operation into GEMM kernel and performs a global synchronization before local reduction. We use a separate stream to perform scatter asynchronously and perform local reduction on another stream. Overall, we achieve $1.28\times$ speedup to PyTorch+NCCL and $1.30\times$ speedup to FLUX.

\textbf{AllGather MoE}

For AG+MoE-intra, FLUX fused MoE kernels focus on expert parallel (EP), while we use tensor parallel (TP). The other baseline PyTorch+NCCL implementation uses Python loops for GroupGEMMs, which is a weak baseline. As a result, our performance is much better compared to this baseline. Overall, the average speedup to PyTorch+NCCL is $44.97\times$. We show the test shapes and our performance in absolute values in Table~\ref{table:ag-moe-shapes}.

\begin{table}[htb]
    \centering
    \footnotesize
    \caption{Test Shapes for AllGather MoE and Performance (ms).}
    \label{table:ag-moe-shapes}
    \begin{tabular}{c|c|c|c|c|c|c|c|c|c}
        \toprule
        \multirow{2}{*}{\textbf{Name}} & \multirow{2}{*}{\textbf{tokens/rank}} & \multirow{2}{*}{\textbf{in hidden}} & \multirow{2}{*}{\textbf{out hidden}} & \multirow{2}{*}{\textbf{experts}} & \multirow{2}{*}{\textbf{topk}} & \multicolumn{2}{c|}{\textbf{Ours}} & \multicolumn{2}{c}{\textbf{PyTorch}}\\
        \cline{7-10}
        & & & & & & \textbf{Intra} & \textbf{Inter} & \textbf{Intra} & \textbf{Inter} \\
        \midrule
        AG+MoE-1 & 256 & 2048 & 1408 & 60 & 4 & 0.33 & 0.45 & 23.95 & 28.84 \\
        \midrule
        AG+MoE-2 & 512 & 2048 & 1408 & 60 & 4 & 0.40 & 1.37 & 26.25 & 29.77 \\
        \midrule
        AG+MoE-3 & 1024 & 2048 & 1408 & 60 & 4 & 0.58 & 1.80 & 30.42 & 43.31 \\
        \midrule
        AG+MoE-4 & 2048 & 2048 & 1408 & 60 & 4 & 0.97 & 3.07 & 55.63 & 63.73 \\
        \midrule
        AG+MoE-5 & 256 & 14336 & 4096 & 8 & 2 & 0.54 & 1.01 & 7.05 & 19.92 \\
        \midrule
        AG+MoE-6 & 512 & 14336 & 4096 & 8 & 2 & 0.72 & 1.89 & 26.34 & 36.07 \\
        \midrule
        AG+MoE-7 & 1024 & 14336 & 4096 & 8 & 2 & 1.19 & 3.41 & 52.99 & 67.61 \\
        \midrule
        AG+MoE-8 & 2048 & 14336 & 4096 & 8 & 2 & 2.10 & 6.51 & 107.32 & 129.30 \\
        \midrule
        AG+MoE-9 & 256 & 16384 & 6144 & 8 & 2 & 0.81 & 1.39 & 11.02 & 27.29 \\
        \midrule
        AG+MoE-10 & 512 & 16384 & 6144 & 8 & 2 & 1.06 & 2.21 & 39.65 & 52.32 \\
        \midrule
        AG+MoE-11 & 1024 & 16384 & 6144 & 8 & 2 & 1.66 & 4.32 & 80.46 & 101.61 \\
        \midrule
        AG+MoE-12 & 2048 & 16384 & 6144 & 8 & 2 & 2.92 & 8.28 & 159.69 & 192.67 \\
        \midrule
        AG+MoE-13 & 512 & 1408 & 2048 & 64 & 6 & 0.45 & 0.84 & 29.25 & 38.17 \\
        \midrule
        AG+MoE-14 & 1024 & 1408 & 2048 & 64 & 6 & 0.67 & 1.26 & 48.86 & 56.77 \\
        \midrule
        AG+MoE-15 & 2048 & 1408 & 2048 & 64 & 6 & 1.18 & 2.18 & 74.26 & 90.44 \\


        \bottomrule
    \end{tabular}
\end{table}

\textbf{MoE ReduceScatter}

For MoE+RS-intra, the overlapping kernels include MoE GroupGEMM, topk reduction, and reduce scatter. Similar to AG-MoE-intra, we show our own performance in absolute values. The performance of the PyTorch+NCCL baseline is much worse to compare. We achieve on average $15.55\times$ speedup.
The test shapes and performance results are shown in Table~\ref{table:moe-rs-shapes}.

\begin{table}[htb]
    \centering
    \footnotesize
    \caption{Test Shapes for MoE ReduceScatter and Performance (ms).}
    \label{table:moe-rs-shapes}
    \begin{tabular}{c|c|c|c|c|c|c|c|c|c}
        \toprule
        \multirow{2}{*}{\textbf{Name}} & \multirow{2}{*}{\textbf{tokens/rank}} & \multirow{2}{*}{\textbf{in hidden}} & \multirow{2}{*}{\textbf{out hidden}} & \multirow{2}{*}{\textbf{experts}} & \multirow{2}{*}{\textbf{topk}} & \multicolumn{2}{c|}{\textbf{Ours}} & \multicolumn{2}{c}{\textbf{PyTorch}}\\
        \cline{7-10}
        & & & & & & \textbf{Intra} & \textbf{Inter} & \textbf{Intra} & \textbf{Inter} \\
\midrule
MoE-RS-1 & 1024 & 1536 & 2048 & 8 & 2 & 0.51 & 3.62 & 4.35 & 12.41 \\
\midrule
MoE-RS-2 & 1024 & 1536 & 2048 & 32 & 2 & 0.55 & 3.90 & 13.89 & 33.05 \\
\midrule
MoE-RS-3 & 1024 & 1536 & 2048 & 64 & 2 & 0.67 & 4.82 & 27.91 & 61.70 \\
\midrule
MoE-RS-4 & 1024 & 1536 & 2048 & 32 & 5 & 0.92 & 7.78 & 14.48 & 35.35 \\
\midrule
MoE-RS-5 & 1024 & 1536 & 2048 & 64 & 5 & 0.93 & 8.25 & 29.96 & 64.88 \\
\midrule
MoE-RS-6 & 1024 & 2048 & 4096 & 8 & 2 & 0.98 & 7.00 & 5.02 & 17.93 \\
\midrule
MoE-RS-7 & 1024 & 2048 & 4096 & 32 & 2 & 1.08 & 7.86 & 14.12 & 38.24 \\
\midrule
MoE-RS-8 & 1024 & 2048 & 4096 & 64 & 2 & 1.34 & 9.87 & 28.61 & 66.48 \\
\midrule
MoE-RS-9 & 1024 & 2048 & 4096 & 32 & 5 & 1.84 & 15.51 & 16.70 & 44.37 \\
\midrule
MoE-RS-10 & 1024 & 2048 & 4096 & 64 & 5 & 1.86 & 16.60 & 27.71 & 71.82 \\


        \bottomrule
    \end{tabular}
\end{table}

\subsection{Inter-node Kernel Performance on Nvidia GPUs}
To scale the overlapping kernels to more nodes, we employ the optimizations introduced in previous sections and show the performance results of different kernels.

\textbf{AllGather GEMM}

For AG+GEMM-inter, we test the performance on 2 nodes of H800 GPUs.
We mainly compare our performance with PyTorch+NCCL, and FLUX using metrics reported in its original paper with some problem size data not available.
The results are shown in Figure~\ref{fig:ag-gemm-inter-node-perf}. We exceed PyTorch+NCCL consistently, with an average speedup of $1.33\times$ and achieve $95.60\%$ the performance of FLUX.

\textbf{GEMM ReduceScatter}

For GEMM+RS-inter, we also use 2 nodes of H800 GPUs for test. We use FLUX's performance reported in the original paper as baseline. We also compare to PyTorch+NCCL. The results are shown in Figure~\ref{fig:gemm-rs-inter-node-perf}. Overall, we can achieve $96.36\%$ the performance of FLUX. Compared to PyTorch+NCCL, the average speedup is $1.42\times$.

\textbf{AllGather MoE and MoE ReduceScatter}

By reusing the inter-node AllGather and ReduceScatter kernels, we also scale MoE kernels to 2 nodes. For AG+MoE-inter, we list the performance in absolute values in Table~\ref{table:ag-moe-shapes}. The results show that scaling AG+MoE kernel from 1 node to 2 nodes increases the latency almost linearly, demonstrating good weak scaling.
However, for MoE+RS-inter, the scaling is not good. As shown in Table~\ref{table:moe-rs-shapes}, when scaling the MoE+RS from 1 node to 2 nodes, the latency increase is not linear. This indicates that a dedicated ReduceScatter kernel is required for MoE+RS to achieve better performance, which is left for future.
Overall, compared to PyTorch+NCCL, the average speedup of AG+MoE-inter is $26.50\times$, the average speedup of MoE+RS-inter is $5.16\times$.

\textbf{Distributed Flash Decoding}

Previous work~\cite{flash-decoding, flashinfer} only implements flash decoding kernel for a single device. We scale flash decoding to more devices, both intra-node and inter-node.
Flash decoding is a bandwidth bound kernel, so the main evaluation metric is the achieved HBM bandwidth for each GPU.
The peak HBM bandwidth for H800 GPU is around 3 TB/s. By gradually increasing the number of GPUs involved in our distributed flash decoding, we can observe the bandwidth change of each GPU.
We use batch size 1 to show the performance of flash decoding.
As shown in Figure~\ref{fig:flash-decode-perf}, when increasing the number of GPUs, if we keep the KV cache length of each GPU unchanged (weak scaling), the achieved HBM bandwidth remains high even for 32 GPUs (1.7 TB/s for 32K KV cache length per GPU).
If we keep the global KV cache length unchanged (strong scaling), the achieved HBM bandwidth decreases as the number of GPUs increases.
The decoding latency indicates that for global KV cache length less than 256K, increasing the number of GPUs is not beneficial. For very long KV cache lengths (e.g. 1M), the more GPUs used, the lower latency can be achieved.
The good scalability of our distributed flash decoding comes from the low-latency AllGather kernel. This distributed flash decoding paves the way to the efficient execution of extremely long context decoding, which might be useful for future reasoning models that are more powerful than existing models.

\textbf{Low-latency AllGather}

Besides NVLink, we also support PCIe communication. To show the performance on PCIe clusters, we use L20 GPUs. L20 only supports PCIe communication. We implement low-latency AllGather for PCIe and test the performance on 8 L20 GPUs and 16 L20 GPUs (2 nodes). The results are shown in Figure~\ref{fig:low-latency-ag-l20}. We compare to NVSHMEM builtin AllGather (fcollect) and NCCL builtin AllGather.
NVSHMEM-32bit refers to using 32bit data type for communication, while NVSHMEM-64bit refers to using 64bit data type for communication. NCCL-inplace and NCCL-out-of-place are different AllGather implementations in NCCL. NCCL-inplace uses the input buffer as output buffer, while NCCL-out-of-place uses another output buffer.
For single node, we achieve the lowest latency on average, achieving bandwidth improvement compared to NVSHMEM ($1.40\times$ to 32bit and $1.48\times$ to 64bit) and NCCL ($3.11\times$ to inplace and $3.00\times$ to out-of-place). 
For 2 nodes, we achieve $1.31\times$ bandwidth improvement to NVSHMEM-64bit and $1.38\times$ to NVSHMEM-32bit.

\textbf{Low-latency AllToAll}

For expert-parallel MoE, AllToAll is mainly used for tokens communication among experts. Previous work such as DeepEP~\cite{deepep} implements extremely high-performance AllToAll kernels for both training and inference. Although providing high performance, the implementation takes thousands lines of CUDA code and is extremely hard to maintain. To demonstrate the advantage of our compiler, we re-implement the AllToAll kernel for inference with only hundreds lines of Python code. We test the inference kernel and scale the number of GPUs from 8 to 64.
The results are shown in Figure~\ref{fig:all2all-perf}. Our Python kernel consistently exceeds DeepEP except for AllToAll Dispatch on 64 GPUs. The average speedup of AllToAll Dispatch is $1.18\times$, while the speedup of AllToAll Combine is $1.44\times$.
Note that although we implement the same functionality as DeepEP, some details are different. First of all, we use IBRC, while DeepEP uses IBGDA. For intra-node communication, we use NVLink, while DeepEP only uses IB. DeepEP implements a much more complicated logic to control memory queue, which is aimed to reduce the waste of GPU memory. But the memory management logic incurs additional overhead. However, we allocate a much larger memory buffer than DeepEP and omit the memory control logic. As a result, we can achieve better performance compared to DeepEP. We also test larger GPU clusters (e.g. 128 GPUs), the results show that DeepEP still produces the best performance. The reason behind this is that DeepEP uses IBGDA, which has better scalability than IBRC. To further improve our performance, we need to use IBGDA to re-implement our kernel. However, current NVSHMEM bitcode library doesn't support IBGDA. So we leave this for future work.

\subsection{Intra-node Kernel Performance on AMD GPUs}

To demonstrate the generality of our compiler, we also show the performance on AMD GPUs. We use MI308X GPUs within one node and use AG+GEMM and GEMM+RS to show the performance.
The results are shown in Figure~\ref{fig:amd-ag-gemm-perf} and Figure~\ref{fig:amd-gemm-rs-perf}. Our baseline is AMD PyTorch+RCCL (PyTorch uses rocBLAS, and rocBLAS provides state-of-the-art GEMM kernels on AMD GPUs). The performance of the code generated by Triton is slightly lower than that of rocBLAS. Even so, we manage to achieve better overlapping performance. For AG+GEMM, we achieve an average speedup of $1.09\times$. For GEMM+RS, we achieve an average speedup of $1.16\times$.

\section{Conclusion}

Distributed programming and accelerator code development have long been challenging problems for many AI infrastructure developers. In the past, the approach relying on low-level programming led to an excessively development difficulty, and the resulting code was extremely difficult to maintain. To address this issue, we propose Triton-distributed. By integrating distributed capabilities into the Triton compiler, we unify the programming languages for distributed kernels and computational kernels. The required development can be completed at the Python level, and in various operators verification experiments, we have demonstrate that the code generation results of our approach can rival hand-written code. This work is fundamental, providing the underlying infrastructure for more developers to engage in distributed development in the future. Our method can be migrated to multiple types of chips.

\clearpage

\bibliographystyle{plainnat}
\bibliography{main}

\begin{thebibliography}{40}
\providecommand{\natexlab}[1]{#1}
\providecommand{\url}[1]{\texttt{#1}}
\expandafter\ifx\csname urlstyle\endcsname\relax
  \providecommand{\doi}[1]{doi: #1}\else
  \providecommand{\doi}{doi: \begingroup \urlstyle{rm}\Url}\fi

\bibitem[AMD(2024)]{rccl}
AMD.
\newblock Rocm communication collectives library.
\newblock \url{https://github.com/ROCm/rccl}, 2024.

\bibitem[Cambricon(2024)]{triton-mlu}
Cambricon.
\newblock Triton-linalg for mlu, 2024.
\newblock URL \url{https://github.com/Cambricon/triton-linalg}.

\bibitem[Chang et~al.(2024)Chang, Bao, Hou, Jiang, Zheng, Zhong, Zhang, Song, Jiang, Lin, Jin, and Liu]{flux}
Li{-}Wen Chang, Wenlei Bao, Qi~Hou, Chengquan Jiang, Ningxin Zheng, Yinmin Zhong, Xuanrun Zhang, Zuquan Song, Ziheng Jiang, Haibin Lin, Xin Jin, and Xin Liu.
\newblock {FLUX:} fast software-based communication overlap on gpus through kernel fusion.
\newblock \emph{CoRR}, abs/2406.06858, 2024.
\newblock \doi{10.48550/ARXIV.2406.06858}.
\newblock URL \url{https://doi.org/10.48550/arXiv.2406.06858}.

\bibitem[Chen et~al.(2024)Chen, Li, Zhu, Duan, Sun, Zhang, and Yang]{centauri}
Chang Chen, Xiuhong Li, Qianchao Zhu, Jiangfei Duan, Peng Sun, Xingcheng Zhang, and Chao Yang.
\newblock Centauri: Enabling efficient scheduling for communication-computation overlap in large model training via communication partitioning.
\newblock In Rajiv Gupta, Nael~B. Abu{-}Ghazaleh, Madan Musuvathi, and Dan Tsafrir, editors, \emph{Proceedings of the 29th {ACM} International Conference on Architectural Support for Programming Languages and Operating Systems, Volume 3, {ASPLOS} 2024, La Jolla, CA, USA, 27 April 2024- 1 May 2024}, pages 178--191. {ACM}, 2024.
\newblock \doi{10.1145/3620666.3651379}.
\newblock URL \url{https://doi.org/10.1145/3620666.3651379}.

\bibitem[Chen et~al.(2018)Chen, Moreau, Jiang, Zheng, Yan, Shen, Cowan, Wang, Hu, Ceze, Guestrin, and Krishnamurthy]{tvm}
Tianqi Chen, Thierry Moreau, Ziheng Jiang, Lianmin Zheng, Eddie~Q. Yan, Haichen Shen, Meghan Cowan, Leyuan Wang, Yuwei Hu, Luis Ceze, Carlos Guestrin, and Arvind Krishnamurthy.
\newblock {TVM:} an automated end-to-end optimizing compiler for deep learning.
\newblock In Andrea~C. Arpaci{-}Dusseau and Geoff Voelker, editors, \emph{13th {USENIX} Symposium on Operating Systems Design and Implementation, {OSDI} 2018, Carlsbad, CA, USA, October 8-10, 2018}, pages 578--594. {USENIX} Association, 2018.

\bibitem[Dao et~al.(2023)Dao, Haziza, Massa, and Sizov]{flash-decoding}
Tri Dao, Daniel Haziza, Francisco Massa, and Grigory Sizov.
\newblock Flash-decoding for long-context inference, 2023.
\newblock URL \url{https://crfm.stanford.edu/2023/10/12/flashdecoding.html}.

\bibitem[DeepSeek{-}AI et~al.(2024)DeepSeek{-}AI, Liu, Feng, Xue, Wang, Wu, Lu, Zhao, Deng, Zhang, Ruan, Dai, Guo, Yang, Chen, Ji, Li, Lin, Dai, Luo, Hao, Chen, Li, Zhang, Bao, Xu, Wang, Zhang, Ding, Xin, Gao, Li, Qu, Cai, Liang, Guo, Ni, Li, Wang, Chen, Chen, Yuan, Qiu, Li, Song, Dong, Hu, Gao, Guan, Huang, Yu, Wang, Zhang, Xu, Xia, Zhao, Wang, Zhang, Li, Wang, Zhang, Zhang, Tang, Li, Tian, Huang, Wang, Zhang, Wang, Zhu, Chen, Du, Chen, Jin, Ge, Zhang, Pan, Wang, Xu, Zhang, Chen, Li, Lu, Zhou, Chen, Wu, Ye, Ye, Ma, Wang, Zhou, Yu, Zhou, Pan, Wang, Yun, Pei, Sun, Xiao, and Zeng]{deepseek-v3}
DeepSeek{-}AI, Aixin Liu, Bei Feng, Bing Xue, Bingxuan Wang, Bochao Wu, Chengda Lu, Chenggang Zhao, Chengqi Deng, Chenyu Zhang, Chong Ruan, Damai Dai, Daya Guo, Dejian Yang, Deli Chen, Dongjie Ji, Erhang Li, Fangyun Lin, Fucong Dai, Fuli Luo, Guangbo Hao, Guanting Chen, Guowei Li, H.~Zhang, Han Bao, Hanwei Xu, Haocheng Wang, Haowei Zhang, Honghui Ding, Huajian Xin, Huazuo Gao, Hui Li, Hui Qu, J.~L. Cai, Jian Liang, Jianzhong Guo, Jiaqi Ni, Jiashi Li, Jiawei Wang, Jin Chen, Jingchang Chen, Jingyang Yuan, Junjie Qiu, Junlong Li, Junxiao Song, Kai Dong, Kai Hu, Kaige Gao, Kang Guan, Kexin Huang, Kuai Yu, Lean Wang, Lecong Zhang, Lei Xu, Leyi Xia, Liang Zhao, Litong Wang, Liyue Zhang, Meng Li, Miaojun Wang, Mingchuan Zhang, Minghua Zhang, Minghui Tang, Mingming Li, Ning Tian, Panpan Huang, Peiyi Wang, Peng Zhang, Qiancheng Wang, Qihao Zhu, Qinyu Chen, Qiushi Du, R.~J. Chen, R.~L. Jin, Ruiqi Ge, Ruisong Zhang, Ruizhe Pan, Runji Wang, Runxin Xu, Ruoyu Zhang, Ruyi Chen, S.~S. Li, Shanghao Lu, Shangyan Zhou,
  Shanhuang Chen, Shaoqing Wu, Shengfeng Ye, Shengfeng Ye, Shirong Ma, Shiyu Wang, Shuang Zhou, Shuiping Yu, Shunfeng Zhou, Shuting Pan, T.~Wang, Tao Yun, Tian Pei, Tianyu Sun, W.~L. Xiao, and Wangding Zeng.
\newblock Deepseek-v3 technical report.
\newblock \emph{CoRR}, abs/2412.19437, 2024.
\newblock \doi{10.48550/ARXIV.2412.19437}.
\newblock URL \url{https://doi.org/10.48550/arXiv.2412.19437}.

\bibitem[Denniston et~al.(2016)Denniston, Kamil, and Amarasinghe]{dist-halide}
Tyler Denniston, Shoaib Kamil, and Saman~P. Amarasinghe.
\newblock Distributed halide.
\newblock In Rafael Asenjo and Tim Harris, editors, \emph{Proceedings of the 21st {ACM} {SIGPLAN} Symposium on Principles and Practice of Parallel Programming, PPoPP 2016, Barcelona, Spain, March 12-16, 2016}, pages 5:1--5:12. {ACM}, 2016.
\newblock \doi{10.1145/2851141.2851157}.
\newblock URL \url{https://doi.org/10.1145/2851141.2851157}.

\bibitem[Feng et~al.(2023)Feng, Hou, Jin, Lin, Shao, Lai, Ye, Zheng, Yu, Yu, and Chen]{tensorir}
Siyuan Feng, Bohan Hou, Hongyi Jin, Wuwei Lin, Junru Shao, Ruihang Lai, Zihao Ye, Lianmin Zheng, Cody~Hao Yu, Yong Yu, and Tianqi Chen.
\newblock Tensorir: An abstraction for automatic tensorized program optimization.
\newblock In Tor~M. Aamodt, Natalie D.~Enright Jerger, and Michael~M. Swift, editors, \emph{Proceedings of the 28th {ACM} International Conference on Architectural Support for Programming Languages and Operating Systems, Volume 2, {ASPLOS} 2023, Vancouver, BC, Canada, March 25-29, 2023}, pages 804--817. {ACM}, 2023.
\newblock \doi{10.1145/3575693.3576933}.
\newblock URL \url{https://doi.org/10.1145/3575693.3576933}.

\bibitem[Google(2025)]{pallas}
Google.
\newblock {Pallas}, 2025.
\newblock URL \url{https://docs.jax.dev/en/latest/pallas/index.html}.

\bibitem[Goumas et~al.(2001)Goumas, Sotiropoulos, and Koziris]{overlap-1}
Georgios~I. Goumas, Aristidis Sotiropoulos, and Nectarios Koziris.
\newblock Minimizing completion time for loop tiling with computation and communication overlapping.
\newblock In \emph{Proceedings of the 15th International Parallel {\&} Distributed Processing Symposium (IPDPS-01), San Francisco, CA, USA, April 23-27, 2001}, page~39. {IEEE} Computer Society, 2001.
\newblock \doi{10.1109/IPDPS.2001.924976}.
\newblock URL \url{https://doi.org/10.1109/IPDPS.2001.924976}.

\bibitem[Huang et~al.(2019)Huang, Cheng, Bapna, Firat, Chen, Chen, Lee, Ngiam, Le, Wu, and Chen]{gpipe}
Yanping Huang, Youlong Cheng, Ankur Bapna, Orhan Firat, Dehao Chen, Mia~Xu Chen, HyoukJoong Lee, Jiquan Ngiam, Quoc~V. Le, Yonghui Wu, and Zhifeng Chen.
\newblock Gpipe: Efficient training of giant neural networks using pipeline parallelism.
\newblock In Hanna~M. Wallach, Hugo Larochelle, Alina Beygelzimer, Florence d'Alch{\'{e}}{-}Buc, Emily~B. Fox, and Roman Garnett, editors, \emph{Advances in Neural Information Processing Systems 32: Annual Conference on Neural Information Processing Systems 2019, NeurIPS 2019, December 8-14, 2019, Vancouver, BC, Canada}, pages 103--112, 2019.
\newblock URL \url{https://proceedings.neurips.cc/paper/2019/hash/093f65e080a295f8076b1c5722a46aa2-Abstract.html}.

\bibitem[Jia et~al.(2019)Jia, Padon, Thomas, Warszawski, Zaharia, and Aiken]{taso}
Zhihao Jia, Oded Padon, James~J. Thomas, Todd Warszawski, Matei Zaharia, and Alex Aiken.
\newblock {TASO:} optimizing deep learning computation with automatic generation of graph substitutions.
\newblock In Tim Brecht and Carey Williamson, editors, \emph{Proceedings of the 27th {ACM} Symposium on Operating Systems Principles, {SOSP} 2019, Huntsville, ON, Canada, October 27-30, 2019}, pages 47--62. {ACM}, 2019.
\newblock \doi{10.1145/3341301.3359630}.
\newblock URL \url{https://doi.org/10.1145/3341301.3359630}.

\bibitem[Kjolstad et~al.(2017)Kjolstad, Kamil, Chou, Lugato, and Amarasinghe]{taco}
Fredrik Kjolstad, Shoaib Kamil, Stephen Chou, David Lugato, and Saman~P. Amarasinghe.
\newblock The tensor algebra compiler.
\newblock \emph{Proc. {ACM} Program. Lang.}, 1\penalty0 ({OOPSLA}):\penalty0 77:1--77:29, 2017.
\newblock \doi{10.1145/3133901}.
\newblock URL \url{https://doi.org/10.1145/3133901}.

\bibitem[Lattner et~al.(2020)Lattner, Pienaar, Amini, Bondhugula, Riddle, Cohen, Shpeisman, Davis, Vasilache, and Zinenko]{mlir}
Chris Lattner, Jacques~A. Pienaar, Mehdi Amini, Uday Bondhugula, River Riddle, Albert Cohen, Tatiana Shpeisman, Andy Davis, Nicolas Vasilache, and Oleksandr Zinenko.
\newblock {MLIR:} {A} compiler infrastructure for the end of moore's law.
\newblock \emph{CoRR}, abs/2002.11054, 2020.
\newblock URL \url{https://arxiv.org/abs/2002.11054}.

\bibitem[Lu et~al.(2015)Lu, Seo, and Balaji]{overlap-2}
Huiwei Lu, Sangmin Seo, and Pavan Balaji.
\newblock {MPI+ULT:} overlapping communication and computation with user-level threads.
\newblock In \emph{17th {IEEE} International Conference on High Performance Computing and Communications, {HPCC} 2015, 7th {IEEE} International Symposium on Cyberspace Safety and Security, {CSS} 2015, and 12th {IEEE} International Conference on Embedded Software and Systems, {ICESS} 2015, New York, NY, USA, August 24-26, 2015}, pages 444--454. {IEEE}, 2015.
\newblock \doi{10.1109/HPCC-CSS-ICESS.2015.82}.
\newblock URL \url{https://doi.org/10.1109/HPCC-CSS-ICESS.2015.82}.

\bibitem[Marjanovic et~al.(2010)Marjanovic, Labarta, Ayguad{\'{e}}, and Valero]{overlap-3}
Vladimir Marjanovic, Jes{\'{u}}s Labarta, Eduard Ayguad{\'{e}}, and Mateo Valero.
\newblock Overlapping communication and computation by using a hybrid mpi/smpss approach.
\newblock In Taisuke Boku, Hiroshi Nakashima, and Avi Mendelson, editors, \emph{Proceedings of the 24th International Conference on Supercomputing, 2010, Tsukuba, Ibaraki, Japan, June 2-4, 2010}, pages 5--16. {ACM}, 2010.
\newblock \doi{10.1145/1810085.1810091}.
\newblock URL \url{https://doi.org/10.1145/1810085.1810091}.

\bibitem[Narayanan et~al.(2021)Narayanan, Shoeybi, Casper, LeGresley, Patwary, Korthikanti, Vainbrand, Kashinkunti, Bernauer, Catanzaro, Phanishayee, and Zaharia]{megatron-lm}
Deepak Narayanan, Mohammad Shoeybi, Jared Casper, Patrick LeGresley, Mostofa Patwary, Vijay Korthikanti, Dmitri Vainbrand, Prethvi Kashinkunti, Julie Bernauer, Bryan Catanzaro, Amar Phanishayee, and Matei Zaharia.
\newblock Efficient large-scale language model training on {GPU} clusters using megatron-lm.
\newblock In Bronis~R. de~Supinski, Mary~W. Hall, and Todd Gamblin, editors, \emph{International Conference for High Performance Computing, Networking, Storage and Analysis, {SC} 2021, St. Louis, Missouri, USA, November 14-19, 2021}, page~58. {ACM}, 2021.
\newblock \doi{10.1145/3458817.3476209}.
\newblock URL \url{https://doi.org/10.1145/3458817.3476209}.

\bibitem[NVIDIA(2022)]{cublas}
NVIDIA.
\newblock {cuBLAS}, 2022.
\newblock URL \url{https://developer.nvidia.com/cublas}.

\bibitem[Nvidia(2022)]{cutlass}
Nvidia.
\newblock Cutlass, 2022.
\newblock URL \url{https://github.com/NVIDIA/cutlass}.

\bibitem[NVIDIA(2024)]{nccl}
NVIDIA.
\newblock Nvidia collective communications library.
\newblock \url{https://developer.nvidia.com/nccl}, 2024.

\bibitem[OpenAI(2022)]{chatgpt}
OpenAI.
\newblock Chatgpt, 2022.
\newblock URL \url{https://chat.openai.com/}.

\bibitem[OpenAI(2024)]{sora}
OpenAI.
\newblock Sora, 2024.
\newblock URL \url{https://openai.com/sora/}.

\bibitem[OpenAI(2025)]{4o-image}
OpenAI.
\newblock Addendum to gpt-4o system card: Native image generation.
\newblock 2025.
\newblock URL \url{https://cdn.openai.com/11998be9-5319-4302-bfbf-1167e093f1fb/Native_Image_Generation_System_Card.pdf}.

\bibitem[Punniyamurthy et~al.(2023)Punniyamurthy, Hamidouche, and Beckmann]{amd-fused}
Kishore Punniyamurthy, Khaled Hamidouche, and Bradford~M Beckmann.
\newblock Optimizing distributed ml communication with fused computation-collective operations.
\newblock \emph{arXiv preprint arXiv:2305.06942}, 2023.

\bibitem[Ragan{-}Kelley et~al.(2013)Ragan{-}Kelley, Barnes, Adams, Paris, Durand, and Amarasinghe]{halide}
Jonathan Ragan{-}Kelley, Connelly Barnes, Andrew Adams, Sylvain Paris, Fr{\'{e}}do Durand, and Saman~P. Amarasinghe.
\newblock Halide: a language and compiler for optimizing parallelism, locality, and recomputation in image processing pipelines.
\newblock In \emph{{ACM} {SIGPLAN} Conference on Programming Language Design and Implementation, {PLDI} '13, Seattle, WA, USA, June 16-19, 2013}, pages 519--530, 2013.
\newblock \doi{10.1145/2491956.2462176}.
\newblock URL \url{https://doi.org/10.1145/2491956.2462176}.

\bibitem[Rasley et~al.(2020)Rasley, Rajbhandari, Ruwase, and He]{deepspeed}
Jeff Rasley, Samyam Rajbhandari, Olatunji Ruwase, and Yuxiong He.
\newblock Deepspeed: System optimizations enable training deep learning models with over 100 billion parameters.
\newblock In Rajesh Gupta, Yan Liu, Jiliang Tang, and B.~Aditya Prakash, editors, \emph{{KDD} '20: The 26th {ACM} {SIGKDD} Conference on Knowledge Discovery and Data Mining, Virtual Event, CA, USA, August 23-27, 2020}, pages 3505--3506. {ACM}, 2020.
\newblock \doi{10.1145/3394486.3406703}.
\newblock URL \url{https://doi.org/10.1145/3394486.3406703}.

\bibitem[Seed(2025)]{doubao-1.5}
ByteDance Seed.
\newblock Doubao-1.5-pro, 2025.
\newblock URL \url{https://team.doubao.com/en/special/doubao_1_5_pro}.

\bibitem[Subramoni et~al.(2017)Subramoni, Chakraborty, and Panda]{overlap-4}
Hari Subramoni, Sourav Chakraborty, and Dhabaleswar~K. Panda.
\newblock Designing dynamic and adaptive {MPI} point-to-point communication protocols for efficient overlap of computation and communication.
\newblock In Julian~M. Kunkel, Rio Yokota, Pavan Balaji, and David~E. Keyes, editors, \emph{High Performance Computing - 32nd International Conference, {ISC} High Performance 2017, Frankfurt, Germany, June 18-22, 2017, Proceedings}, volume 10266 of \emph{Lecture Notes in Computer Science}, pages 334--354. Springer, 2017.
\newblock \doi{10.1007/978-3-319-58667-0\_18}.
\newblock URL \url{https://doi.org/10.1007/978-3-319-58667-0\_18}.

\bibitem[Team(2024)]{qwen-max}
Qwen Team.
\newblock Qwen2.5 technical report.
\newblock \emph{arXiv preprint arXiv:2412.15115}, 2024.

\bibitem[Team(2025)]{tilelang}
TileLang Team.
\newblock Tilelang, 2025.
\newblock URL \url{https://github.com/tile-ai/tilelang}.

\bibitem[Tillet et~al.(2019)Tillet, Kung, and Cox]{triton}
Philippe Tillet, Hsiang{-}Tsung Kung, and David~D. Cox.
\newblock Triton: an intermediate language and compiler for tiled neural network computations.
\newblock In Tim Mattson, Abdullah Muzahid, and Armando Solar{-}Lezama, editors, \emph{Proceedings of the 3rd {ACM} {SIGPLAN} International Workshop on Machine Learning and Programming Languages, MAPL@PLDI 2019, Phoenix, AZ, USA, June 22, 2019}, pages 10--19. {ACM}, 2019.
\newblock \doi{10.1145/3315508.3329973}.
\newblock URL \url{https://doi.org/10.1145/3315508.3329973}.

\bibitem[Yadav et~al.(2022)Yadav, Aiken, and Kjolstad]{distal}
Rohan Yadav, Alex Aiken, and Fredrik Kjolstad.
\newblock {DISTAL:} the distributed tensor algebra compiler.
\newblock In Ranjit Jhala and Isil Dillig, editors, \emph{{PLDI} '22: 43rd {ACM} {SIGPLAN} International Conference on Programming Language Design and Implementation, San Diego, CA, USA, June 13 - 17, 2022}, pages 286--300. {ACM}, 2022.
\newblock \doi{10.1145/3519939.3523437}.
\newblock URL \url{https://doi.org/10.1145/3519939.3523437}.

\bibitem[Ye et~al.(2025)Ye, Chen, Lai, Lin, Zhang, Wang, Chen, Kasikci, Grover, Krishnamurthy, and Ceze]{flashinfer}
Zihao Ye, Lequn Chen, Ruihang Lai, Wuwei Lin, Yineng Zhang, Stephanie Wang, Tianqi Chen, Baris Kasikci, Vinod Grover, Arvind Krishnamurthy, and Luis Ceze.
\newblock Flashinfer: Efficient and customizable attention engine for llm inference serving.
\newblock \emph{arXiv preprint arXiv:2501.01005}, 2025.
\newblock URL \url{https://arxiv.org/abs/2501.01005}.

\bibitem[Zhang et~al.(2025)Zhang, Zheng, Lin, Jiang, Bao, Jiang, Hou, Cui, Zheng, Chang, Chen, and Liu]{comet}
Shulai Zhang, Ningxin Zheng, Haibin Lin, Ziheng Jiang, Wenlei Bao, Chengquan Jiang, Qi~Hou, Weihao Cui, Size Zheng, Li{-}Wen Chang, Quan Chen, and Xin Liu.
\newblock Comet: Fine-grained computation-communication overlapping for mixture-of-experts.
\newblock \emph{CoRR}, abs/2502.19811, 2025.
\newblock \doi{10.48550/ARXIV.2502.19811}.
\newblock URL \url{https://doi.org/10.48550/arXiv.2502.19811}.

\bibitem[Zhao et~al.(2025)Zhao, Zhou, Zhang, Deng, Xu, Liu, Yu, Li, and Zhao]{deepep}
Chenggang Zhao, Shangyan Zhou, Liyue Zhang, Chengqi Deng, Zhean Xu, Yuxuan Liu, Kuai Yu, Jiashi Li, and Liang Zhao.
\newblock Deepep: an efficient expert-parallel communication library.
\newblock \url{https://github.com/deepseek-ai/DeepEP}, 2025.

\bibitem[Zheng et~al.(2020{\natexlab{a}})Zheng, Jia, Sun, Wu, Yu, Haj{-}Ali, Wang, Yang, Zhuo, Sen, Gonzalez, and Stoica]{ansor}
Lianmin Zheng, Chengfan Jia, Minmin Sun, Zhao Wu, Cody~Hao Yu, Ameer Haj{-}Ali, Yida Wang, Jun Yang, Danyang Zhuo, Koushik Sen, Joseph~E. Gonzalez, and Ion Stoica.
\newblock Ansor: Generating high-performance tensor programs for deep learning.
\newblock In \emph{14th {USENIX} Symposium on Operating Systems Design and Implementation, {OSDI} 2020, Virtual Event, November 4-6, 2020}, pages 863--879. {USENIX} Association, 2020{\natexlab{a}}.
\newblock URL \url{https://www.usenix.org/conference/osdi20/presentation/zheng}.

\bibitem[Zheng et~al.(2020{\natexlab{b}})Zheng, Liang, Wang, Chen, and Sheng]{flextensor}
Size Zheng, Yun Liang, Shuo Wang, Renze Chen, and Kaiwen Sheng.
\newblock Flextensor: An automatic schedule exploration and optimization framework for tensor computation on heterogeneous system.
\newblock In James~R. Larus, Luis Ceze, and Karin Strauss, editors, \emph{{ASPLOS} '20: Architectural Support for Programming Languages and Operating Systems, Lausanne, Switzerland, March 16-20, 2020 {[ASPLOS} 2020 was canceled because of {COVID-19]}}, pages 859--873. {ACM}, 2020{\natexlab{b}}.
\newblock \doi{10.1145/3373376.3378508}.
\newblock URL \url{https://doi.org/10.1145/3373376.3378508}.

\bibitem[Zheng et~al.(2022)Zheng, Chen, Wei, Jin, Han, Lu, Wu, Li, Yan, and Liang]{amos}
Size Zheng, Renze Chen, Anjiang Wei, Yicheng Jin, Qin Han, Liqiang Lu, Bingyang Wu, Xiuhong Li, Shengen Yan, and Yun Liang.
\newblock {AMOS:} enabling automatic mapping for tensor computations on spatial accelerators with hardware abstraction.
\newblock In Valentina Salapura, Mohamed Zahran, Fred Chong, and Lingjia Tang, editors, \emph{{ISCA} '22: The 49th Annual International Symposium on Computer Architecture, New York, New York, USA, June 18 - 22, 2022}, pages 874--887. {ACM}, 2022.
\newblock \doi{10.1145/3470496.3527440}.
\newblock URL \url{https://doi.org/10.1145/3470496.3527440}.

\bibitem[Zheng et~al.(2025)Zheng, Fang, Zheng, Hou, Bao, Zheng, Jiang, Wang, Ye, Lin, Chang, and Liu]{tilelink}
Size Zheng, Jin Fang, Xuegui Zheng, Qi~Hou, Wenlei Bao, Ningxin Zheng, Ziheng Jiang, Dongyang Wang, Jianxi Ye, Haibin Lin, Li-Wen Chang, and Xin Liu.
\newblock Tilelink: Generating efficient compute-communication overlapping kernels using tile-centric primitives, 2025.
\newblock URL \url{https://arxiv.org/abs/2503.20313}.

\end{thebibliography}




\end{document}